\documentclass[11pt,a4paper]{article}
\usepackage{jcappub}
\usepackage{graphicx,epsfig}
\usepackage{amsmath,amssymb}
\usepackage{xcolor}
\usepackage{lineno,hyperref}
\usepackage{wrapfig}
\usepackage{subfigure}
\modulolinenumbers[5]

\renewcommand{\vec}[1]{\ensuremath{\mathbf{#1}}} 
\newcommand{\ii}{\mathrm{i}} 
 
\renewcommand{\phi}{\varphi}

\newcommand{\bfxi}{\mbox{\boldmath{$\xi$}}}
\newcommand{\bfeta}{\mbox{\boldmath{$\eta$}}}
\newcommand{\bfth}{\mbox{\boldmath{$\theta$}}}

\newcommand{\beq}{\begin{equation}}
\newcommand{\beqa}{\begin{eqnarray}}
\newcommand{\eeq}{\end{equation}}
\newcommand{\eeqa}{\end{eqnarray}}

\begin{document}

\title{Lensing of gravitational waves: universal signatures in the beating pattern}

\author{Oleg Bulashenko}
\author{and Helena Ubach}

\affiliation{Departament de F\'{i}sica Qu\`{a}ntica i Astrof\'{i}sica, Institut de Ci\`{e}ncies del Cosmos (ICCUB), \\
Facultat de F\'{i}sica, Universitat de Barcelona, Mart\'{i} i Franqu\`{e}s 1,
E-08028 Barcelona, Spain.}

\emailAdd{oleg@fqa.ub.edu}
\emailAdd{helenaubach@icc.ub.edu}


\abstract{
When gravitational waves propagate near massive objects, their paths curve resulting in gravitational lensing, which is expected to be a promising new instrument in astrophysics. 
If the time delay between different paths is comparable with the wave period, lensing may induce beating patterns in the waveform, and it is very close to caustics that these effects are likely to be observable.
Near the caustic, however, the short-wave asymptotics associated with the geometrical optics approximation breaks down. In order to describe properly the crossover from wave optics to geometrical optics regimes, along with the Fresnel number, which is the ratio between the Schwarzschild diameter of the lens and the wavelength, one has to include another parameter - namely, the angular position of the source with respect to the caustic.
By considering the point mass lens model, we show that in the two-dimensional parameter space, the nodal and antinodal lines for the transmission factor closely follow hyperbolas in a wide range of values near the caustic.
This allows us to suggest a simple formula for the onset of geometrical-optics oscillations which relates the Fresnel number with the angular position of the source in units of the Einstein angle.
We find that the mass of the lens can be inferred from the analysis of the interference fringes of a specific lensed waveform.
}


\maketitle
\flushbottom

{
\hypersetup{linkcolor=blue}
\tableofcontents
}

\section{Introduction}

When an electromagnetic or gravitational wave emitted by a distant cosmological 
source travels across the Universe to reach our detectors, its path is 
perturbed by a gravitational field, i.e., by the geometry of spacetime. This is 
called gravitational lensing---the phenomenon first attributed to the bending of 
light due to gravity. Nowadays, this is regarded as an indispensable tool in 
astrophysics, with applications ranging from detecting exoplanets to constraining 
the distribution of dark matter and determining cosmological parameters
\cite{schneider-92, petters-01}.

To date, only the lensing of electromagnetic (EM) waves has been confidently 
observed. 
However, since the first direct detection of gravitational waves (GWs) in 2015 
\cite{LIGO16-1}, and the subsequent registration of about $90$ new GW events 
by LIGO and Virgo \cite{LIGO18-GWTC1, LIGO20-GWTC2, LIGO21-GWTC3},
the gravitational lensing of GWs has become the focus of investigation (see recent reviews 
\cite{meena-20,mukherjee-20} and references therein).
As the sensitivities of GW detectors improve, and upcoming new detector 
facilities join the network in the near future (with a wide-range frequency scale 
from nano-Hz to kHz \cite{auger-plagnol-17, maggiore-20, barrause-20, bailes-21,
Corbin:2005ny,Ruan:2018tsw,AEDGE:2019nxb,Badurina:2019hst,
Kawamura:2020pcg,TianQin:2020hid})
the gravitational lensing of GWs is expected to be a promising new instrument 
in astrophysics.

Gravitational lensing is manifested in different ways, depending on the mass and the nature of the lens system
\cite{deguchi86a,deguchi86b}.
For massive lenses (galaxies or galaxy clusters), one would expect either magnification 
\cite{wang96,dai17,Ng:2017yiu,lai18,Broadhurst:2018saj,Hannuksela:2019kle} or multiple images 
of GWs \cite{sereno10,takahashi17,smith18,li18,oguri19,liu19,ezquiaga-holz-21} in the frequency band of LIGO/Virgo.
The latter case, also called {\it strong lensing}, means the detection of repeated events separated by a time delay.
For more compact lenses, when the Schwarzschild radius $R_{\rm S}$ is comparable to the characteristic wavelength $\lambda$ of a GW,  
the interference between the images can produce beating patterns in the waveform.
This effect,  also called {\it microlensing}, has recently attracted much attention
\cite{nakamura98,nakamura99,takahashi03,matsunaga06,cao14,Christian:2018vsi,dai18,jung19,liao19,
diego19,hou20,orazio20,Liao:2020hnx,hou21,cremonese21,cremonese21b,yu21,
Wang:2021lij,urrutia21,biesiada21,chung21,suvorov21,yeung21,dalang22,gais22,ramesh22,basak22,gao22}.

In this paper, we address the problem of gravitational lensing from a general full-wave optics perspective \cite{born-wolf-03}
paying special attention to the conditions under which the interference between the virtual images is essential (microlensing).
With the aim of finding some universal signatures in the interference pattern appropriate for the creation of templates for interferometric measurements, we consider the most generic lens model---the point mass lens (PML), also called Schwarzschild lens.
This is suitable for lens objects like stars, black holes (BHs), compact dark matter clumps, etc.,  whose dimensions are much smaller than the Einstein radius---the relevant lengthscale in gravitational lensing.
Despite the fact that the PML model has been widely used for both EM \cite{schneider-92,petters-01} and GW lensing 
\cite{nakamura98,nakamura99,takahashi03,matsunaga06,cao14,Christian:2018vsi,jung19,liao19,
diego19,hou20,orazio20,Liao:2020hnx,cremonese21,cremonese21b,yu21,
Wang:2021lij,urrutia21,biesiada21,chung21,suvorov21,yeung21,dalang22,gais22,ramesh22,basak22},
some issues still need to be clarified.
In particular, the customary condition for the transition from wave optics to geometrical optics (GO) regimes
\cite{nakamura98,takahashi03,matsunaga06,oguri19}, $\lambda\ll R_{\rm S}$,  breaks down near the caustic 
\cite{nakamura99, ezquiaga-holz-21,berry21,choi21}.
Indeed, when the source approaches the line of sight (a caustic point for the PML), the time delay between the images becomes infinitesimally small, which means one needs infinite frequencies to reach the GO limit.

In order to describe properly this limit, 
we analyze in detail the transition from full-wave to GO regimes in a two-dimensional space of characteristic parameters---the Fresnel number $\nu=2R_{\rm S}/\lambda$ (the frequency parameter) and 
the angular position of the source $y$ in units of the Einstein angle (the alignment parameter).
By analyzing the transmission factor in the $(\nu,y)$ space, we distinguish three regions which are of interest: diffraction, amplification and geometrical-optics oscillations. We suggest analytical formulas which separate the regions. 
In particular, for the onset of the GO oscillations we suggest the following condition: the time delay between the images should fulfill $f\Delta t_{21}>1/2$, with $f$ being the characteristic frequency of a GW. 
This condition is less restrictive than $f\Delta t_{21}\gg 1$, which is usually considered for the GO limit \cite{takahashi03}. 
We also show that under the condition which we call {\it "close alignment"} ($y\lesssim 0.5$), 
the time delay function can be approximated by $\tau_{21}\approx 2y$, which gives us a simple formula for the GO onset, $\nu y > 1/4$. Translated into physical units, it can be used for quick estimates of the lower bound on  the lens mass $M$ above which the GO approximation holds. Namely,
\begin{equation}
\frac{M}{M_{\odot}} >  1.25 \times 10^4 \left(\frac{f}{\rm Hz}\right)^{-1} \left(\frac{1}{y}\right).
\label{eq:GO-limit}
\end{equation}
In contrast to previous studies, Eq.~\eqref{eq:GO-limit} along with the frequency $f$ includes the alignment parameter $y$ and, written in this way, it is also valid close to the caustic.
For larger $y$, as will be shown later, the condition \eqref{eq:GO-limit} can be generalized by including the next-order terms in the expansion of $\tau_{21}(y)$. Alternatively, one can use numerical adjustment as in Ref.~\cite{urrutia21}.

The presented analysis is not restricted to gravitational lensing of GWs,
it can also be applied to the lensing of EM waves or any scalar waves in the PML geometry. 
Recently, it was pointed out that ``diffractive gravitational lensing'' can be 
used to probe as yet undiscovered astrophysical objects like primordial BHs or 
ultra-compact dark matter minihalos, made up for instance of QCD axions. 
In this investigation, different frequency ranges of EM are explored: from 
femtolensing of gamma ray bursts (GRBs)  \cite{katz18,jung20,paynter21} 
and optical microlensing \cite{montero-camacho19,sugiyama20} to radiowaves of fast 
radio bursts (FRBs) \cite{katz20,jow20}. 
It should be noted, however, that for GWs, due to their longer wavelengths and 
the coherence preserved over cosmological distances, the wave optics effects are 
more substantial than those for EM waves.
In the latter case, the interference can also be washed out by the incoherent 
nature of EM waves emitted from different parts of an extended source \cite{matsunaga06,katz18}.

This paper is organized as follows. 
In Sec.~\ref{sec-lensing}, we review the Fresnel-Kirchhoff approach for the
diffraction of scalar waves on a thin gravitational lens and introduce the 
characteristic parameters for the time and frequency scales.
To characterize the relevance of wave optics effects, we define a two-dimensional phase 
function at the lens plane (phase screen) and introduce the Fresnel number 
$\nu$ as a crucial parameter for the analysis of the contribution of partial waves 
to the transmission factor.
The geometrical optics limit is discussed in Sec.~\ref{sec-GO}. 
In Sec.~\ref{Sec_full-GO} we provide a detailed analysis of the transmission 
factor and demonstrate that the lines of local maxima and minima in the 
interference pattern can be regarded as constant-phase lines between two GO 
rays and they can be approximated by hyperbolas in a wide range of values near the caustic.
The nodal and antinodal lines, when the source position is fixed, lead to the oscillation pattern uniformly spaced over the frequency, and similarly, when the frequency of the wave is fixed, the oscillations are uniformly spaced over the source position.
The topological (Morse) phase shift between the images can be extracted from the width of the central maximum.
We also show that the full-wave and the GO approximation start practically to 
coincide when the ``optical path difference'' between the GO rays is equal to 
$\lambda/2$.
This allows us to suggest a simple formula for the onset of geometrical-optics oscillations which relates the Fresnel number with the angular position of the source in units of the Einstein angle.
Finally, in Sec.~\ref{sect-waveform} we study the effect of gravitational lensing on the ringdown waveform.
The interference between the images is manifested as beating fringes in the frequency domain of the lensed waveform. From the analysis of the fringes one can infer the mass of the lens.
The conclusions are summarized in  Sec.~\ref{sec-concl}.

\section{Wave approach to gravitational lensing}
\label{sec-lensing}

In this section, we review some aspects of wave effects in gravitational 
lensing, which are relevant for our analysis. 
Let us consider the gravitational waves propagating under the gravitational 
potential of the lens object. 
For the moment, to make the derivation more transparent, we do not include the cosmological expansion in the metric. 
The Minkowskian spacetime disturbed by the lens is given by
\cite{schneider-92}
\begin{equation}
  ds^2 = - \left( 1+\frac{2U}{c^2} \right) c^2 dt^2 + \left( 1- \frac{2U}{c^2} \right) d \mathbf{r}^2,
\label{metric}  
\end{equation}
where $U(\mathbf{r})$ is the gravitational potential of the lens ($|U|/c^2\ll 1$). 
The propagation of electromagnetic as well as gravitational waves (in an 
appropriately chosen gauge) can be described to the leading order by a scalar 
wave equation when the effect of lensing on polarization is 
negligible \cite{peters74, misner73}.
Decomposing the scalar field into Fourier modes
\begin{equation}
\phi(\vec{r},t) = \int_{-\infty}^\infty \widetilde{\phi}(\vec{r},\omega)\,
e^{-\ii \omega t} \,\frac{d\omega}{2\pi},
\end{equation}
one obtains the wave equation for the scalar amplitude 
$\widetilde{\phi}(\vec{r},\omega)$ in the frequency domain \cite{deguchi86a}
\begin{equation}
\left(\nabla^2 + \frac{\omega^2}{c^2} \right)\widetilde{\phi}= \frac{4\,\omega^2}{c^4} \, U\widetilde{\phi},
\label{scalar_wave}
\end{equation}
which is known in wave-optics physics as
an inhomogeneous Helmholtz equation.

\subsection{Thin-lens approximation}

\begin{figure}[t]
\centering
\includegraphics[width=0.7\textwidth]{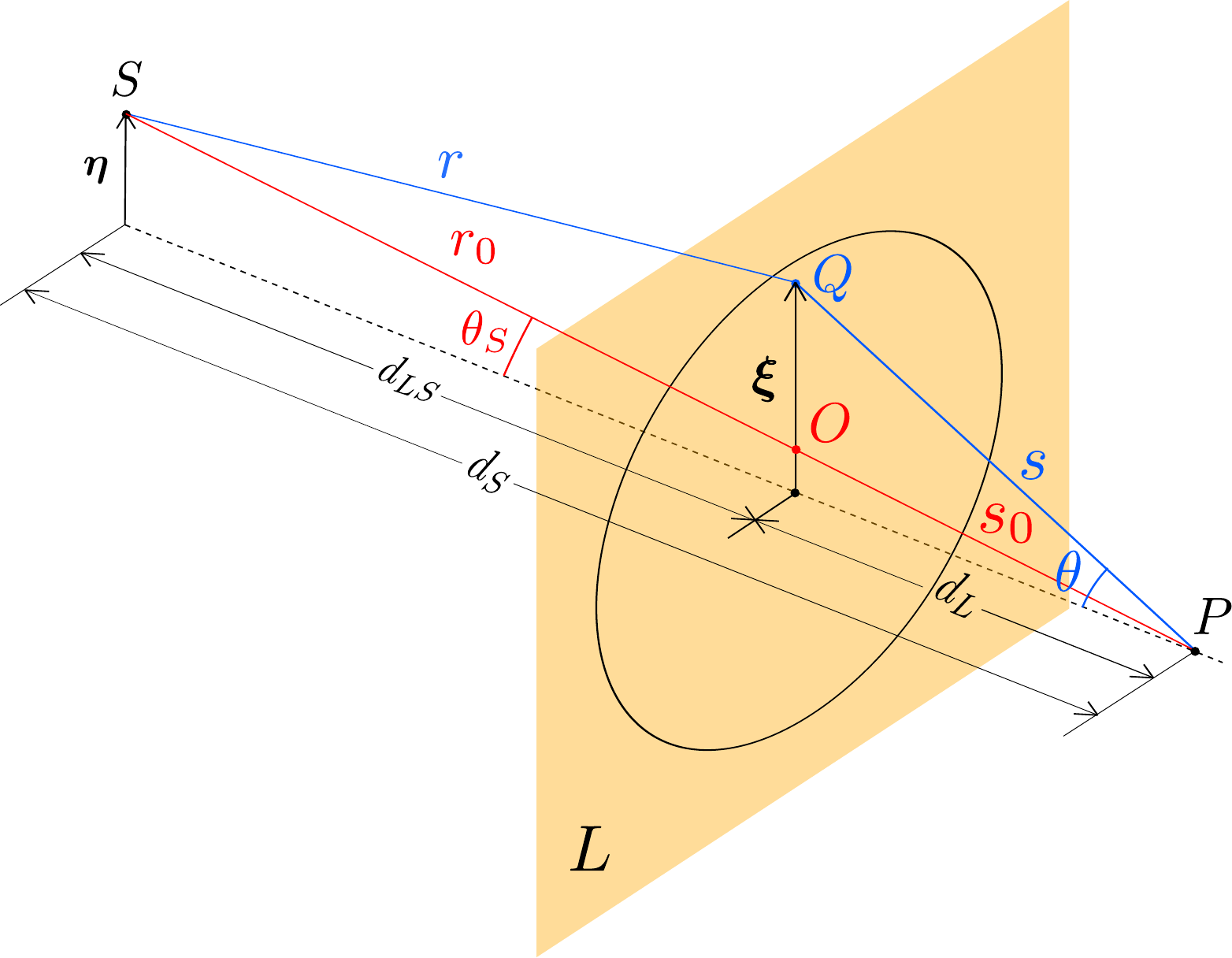}
\caption{Schematic diagram of gravitational lensing in a thin-lens approximation. 
The lens plane $L$ is orthogonal to the optical axis (shown by dashes).
A partial wave (shown in blue) emitted by the source $S$ is scattered at the 
point $Q$ of the lens before it arrives to the observer at $P$. 
In the absence of the lens, this wave would follow the unperturbed path $SOP$ shown in red.
Other notations are explained in the text.}
\label{fig:geom}
\end{figure}
The gravitational lens geometry is depicted in Fig.~\ref{fig:geom}, where
$d_L$, $d_S$, and $d_{LS}$ are the distances of the lens and the source from 
the observer, and of the source from the lens, respectively. 
In a thin-lens approximation, the lensing occurs in a relatively small region 
as compared to the cosmological distances travelled by the waves, $d_L$, 
$d_{LS}$, $d_S$. 
Under this approximation, the lens mass is projected onto the lens plane $L$. 
Hence, the waves are assumed to propagate freely outside the lens and 
interact with a two-dimensional gravitational potential at the lens plane where 
the trajectory is suddenly deflected \cite{schneider-92}. 
Figure~\ref{fig:geom} shows the path of one of the partial waves (blue 
line) emitted by the source $S$,  which is deflected at the point $Q$ of the 
lens and then arrives to the observer at $P$. 
The apparent angle of arrival $\theta$ differs from the real angular location 
of the source $\theta_S$ due to the effect of lensing. 
Then, in the full-wave optics approach, we must take into account all the 
partial waves emitted by the source, which impact the lens plane at different 
places, but arrive at the same point $P$ where the total wave field is detected.

\subsection{Fresnel-Kirchhoff diffraction}
The basic idea of the Huygens-Fresnel theory is that the wave field at the 
observation point $P$ arises from the superposition of secondary waves that 
proceed from the lens surface (see Fig.~\ref{fig:geom}). 
Provided that the radius of curvature of the wavefront at the lens plane is large 
compared to the wavelength, and the angles involved are small, the wave field 
at the observer $P$ can be written in the form of the Fresnel-Kirchhoff 
diffraction integral over the surface of the lens \cite{born-wolf-03}

\begin{equation}
\widetilde{\phi}(P)=\frac{A}{\ii\lambda} \iint \frac{1}{rs} 
~e^{\ii [k(r+s)+\psi_L] } \,dS ,
\label{FK}
\end{equation}
where $A$ is the scalar amplitude of the emitted wave, $\lambda=2\pi/k$ is 
the wavelength, $r$ and $s$ are the distances indicated in Fig.~\ref{fig:geom},
and $\psi_L$ is the gravitational phase shift due to the lensing 
potential which will be defined later on.
It is seen, that the lens acts on the wave as a transparent phase screen: 
passing through the lens, each partial wave acquires the phase shift $\psi_L$ which 
depends 
on the impact parameter and therefore varies over the surface. The wavefront 
after the lens is disturbed due to this phase shift $\psi_L$, hence the 
diffraction 
effects appear at the observation point. 
As Fresnel pointed out in his classical work \cite{fresnel1821,crew1900}, in 
order to produce the phenomena of diffraction ``all that is required is that a 
part of the wave should be retarded with respect to its neighbouring parts.''
This is precisely what happens in the lensing effect.

As the element $dS$ explores the domain of integration in the integral 
(\ref{FK}), the factor $1/(rs)$ is the slowly changing function (due to $r,s \gg 
\lambda$), and it can be replaced by $1/(r_0 s_0)\approx 1/(d_L d_{LS})$ and 
taken away of the integral. The rapidly oscillating exponent, in contrast, 
should be calculated more carefully. It is convenient to add and subtract the 
unlensed path $r_0+s_0$ in the phase to obtain

\begin{equation}
\widetilde{\phi}(P)=\frac{A}{\ii\lambda} \frac{e^{\ii k(r_0+s_0)}}{d_L d_{LS}}
 \iint  e^{\ii [k \, \Delta l+\psi_L]} \, dS
\end{equation}
where $\Delta l=r+s-r_0-s_0$ is the geometrical path difference between the 
lensed and unlensed partial waves. 
The unlensed wave field at the point $P$ can be written as
\begin{equation}
\widetilde{\phi}_0(P)=\frac{A}{d_S} e^{\ii k(r_0+s_0)}
\end{equation}
where, due to $d_S \gg \lambda$, we replaced $1/(r_0+s_0)$ by $1/d_S$ in the 
pre-exponential factor. 

It is convenient to define the transmission factor (which is called the 
transmission function in Ref.~\cite{born-wolf-03} and the amplification factor 
in Ref.~\cite{takahashi03}) as the ratio between the lensed and unlensed GW 
amplitudes at the point $P$
\begin{equation}
F=\frac{\widetilde{\phi}(P)}{\widetilde{\phi}_0(P)} = 
\frac{1}{\ii\lambda} ~ \frac{d_S}{d_L d_{LS}}
 \iint  e^{\ii [k \, \Delta l+\psi_L]} \, dS.
\label{F_transm}
\end{equation}
Following Ref.~\cite{schneider-92}, the geometrical path difference $\Delta l$ 
can be expressed to the leading order (Fresnel expansion) through the source 
position vector $\bfeta$ in the source plane (see Fig.~\ref{fig:geom}) and the 
location of a running vector $\bfxi$ on the lens plane 
\begin{equation}
\Delta l \,(\bfxi,\bfeta) = \frac{d_S}{2 d_L d_{LS}}
 \left( \bfxi - \frac{d_L}{d_S}\,\bfeta \right)^2,
\label{delta_l}
\end{equation}
whereas the gravitational phase shift is given by
\begin{equation}
\psi_L (\bfxi) = - \frac{4Gk}{c^2} \int \Sigma(\bfxi')\,\ln |\bfxi-\bfxi'|\, d^2 
\bfxi'
\label{psi}
\end{equation}
where $\Sigma$ is the surface mass density of the lens. 
For the point mass, which is a prototype of compact lens objects, it is given by 
$\Sigma=M \,\delta^2(\bfxi')$, where $M$ is the mass of the lens.

If no gravitational lens is present on the pathway from the source to the 
observer, i.e.~$\psi_L=0$, the geometrical path difference $\Delta l$ is the 
only function which affects the phases of partial waves. 
For this case, it can be verified, that the integral \eqref{F_transm} gives 
obviously $|F|=1$, which is in accordance with the Huygens-Fresnel principle and 
the definition of the 
transmission factor.

\subsection{Characteristic scales and dimensionless parameters}

Lensing effects are expected to be significant only when the lens is located 
very close to the line of sight, i.e., the source, lens, and observer are 
all aligned within approximately the Einstein angle, 
$\theta_{\rm E}=R_{\rm E}/d_L$, where 
\begin{equation}
R_{\rm E} = \sqrt{2R_{\rm S} \frac{d_L d_{LS}}{d_S}}
\label{R_E}
\end{equation}
is the Einstein radius
and $R_{\rm S}=2GM/c^2$ is the Schwarz\-schild radius of the lens of mass $M$.
Accordingly, it is convenient to normalize the angles $\theta$ and $\theta_S$  
(see Fig.~\ref{fig:geom}) by the Einstein angle $\theta_{\rm E}$ and introduce 
dimensionless vectors $\mathbf{x}$ and $\mathbf{y}$ which determine the location 
of the running vector and the position of the source, respectively, at the lens 
plane, as follows \cite{schneider-92}:
\beq
\mathbf{x}= 
\frac{\bfth}{\theta_{\rm E}} = 
\frac{\bfxi}{R_{\rm E}}, 
\qquad \mathbf{y} = 
\frac{\bfth_S}{\theta_{\rm E}} = 
\frac{d_L}{d_S} \frac{\bfeta}{R_{\rm E}}.
\label{xy}
\eeq

After rescaling \eqref{xy}, the phase function in the exponent of the integral 
(\ref{F_transm}) takes on the form
\begin{equation}
k\Delta l + \psi_L = \omega\, \frac{2R_{\rm S}}{c}
\left[ \frac{1}{2} ( {\bf x} - {\bf y} )^2 - \psi ({\bf x}) + \psi_0 \right],
\label{delta_ln}
\end{equation}
where 
$\psi=-\psi_L/(2R_{\rm S}k)$ is the scaled phase shift due to the lensing potential.
For the PML it is simply $\psi(\mathbf{x})=\ln{|\mathbf{x}|}$.
The phase is defined up to an arbitrary constant $\psi_0$ which does not alter 
the absolute value of the transmission factor $|F|$ we are interested in.

It is seen from Eq.~\eqref{delta_ln} that the time scale is 
determined by 
\begin{equation}
t_M =\frac{2R_{\rm S}}{c},
\label{eq:t_M}
\end{equation}
which is the crossing time of the Schwarzschild diameter.
Thus, to the leading order, the time delay depends only on the mass of the lens 
and is practically independent of the distances either to the source or to the 
lens:
\begin{equation}
t_M \simeq 2 \times 10^{-5}\, {\rm s} 
\,\left(\frac{M}{M_\odot}\right).
\label{eq:t_M2}
\end{equation}
Introducing the dimensionless frequency
\begin{equation}
\nu = \left(\frac{\omega}{2\pi} \right)\, t_M,
\label{eq:nu}
\end{equation}
the transmission factor \eqref{F_transm} can finally be written as
\begin{equation}
F = -\ii \, \nu
\iint e^{2\pi \ii \,\nu\, \tau(\mathbf{x},\mathbf{y})} \, d^2\mathbf{x}
\label{F_nu}
\end{equation}
in which the dimensionless scalar function
\beqa
\tau(\mathbf{x},\mathbf{y}) = \frac{1}{2} ( \mathbf{x}-\mathbf{y} )^2
- \psi(\mathbf{x}) + \psi_0
\label{Fermat-pot}
\eeqa 
can be associated with the Fermat potential \cite{schneider-92} (also called 
time-delay function \cite{takahashi03}).
It represents retardation of partial waves to arrive at the observation point 
due to either the geometrical time delay (the first term) or gravitational one
(the second term). 
Again, for $\psi(\mathbf{x})=0$ (no lensing), the transmission factor \eqref{F_nu} is just $|F|=1$.

It should be noted that the cosmological expansion can also be included in the metric \eqref{metric} \cite{schneider-92}.
The distances $d_L$, $d_S$, and $d_{LS}$ should then be interpreted as angular-diameter distances, for which in general 
$d_S \neq d_L + d_{LS}$ (see Ref.~\cite{schneider-92} for the details). 
This leads to the modifications in the transmission factor \eqref{F_transm}: the integral and the phase in the exponential are multiplied by the factor $(1+z_L)$, where $z_L$ is the redshift of the lens, (see, e.g.,  \cite{takahashi03,matsunaga06,urrutia21,gao22}).
This is equivalent to rescale the frequency $\omega$ of a GW, as $\omega (1+z_L)$ \cite{takahashi03,matsunaga06}.
Finally, one can use the same set of equations  \eqref{eq:t_M}--\eqref{Fermat-pot}, but with the lens mass $M$ replaced by its redshifted value $M_{zL}\equiv M(1+z_L)$.

\subsection{Phase function. Fresnel number}

\begin{figure*}[t]
\centering
\includegraphics[width=0.32\columnwidth]{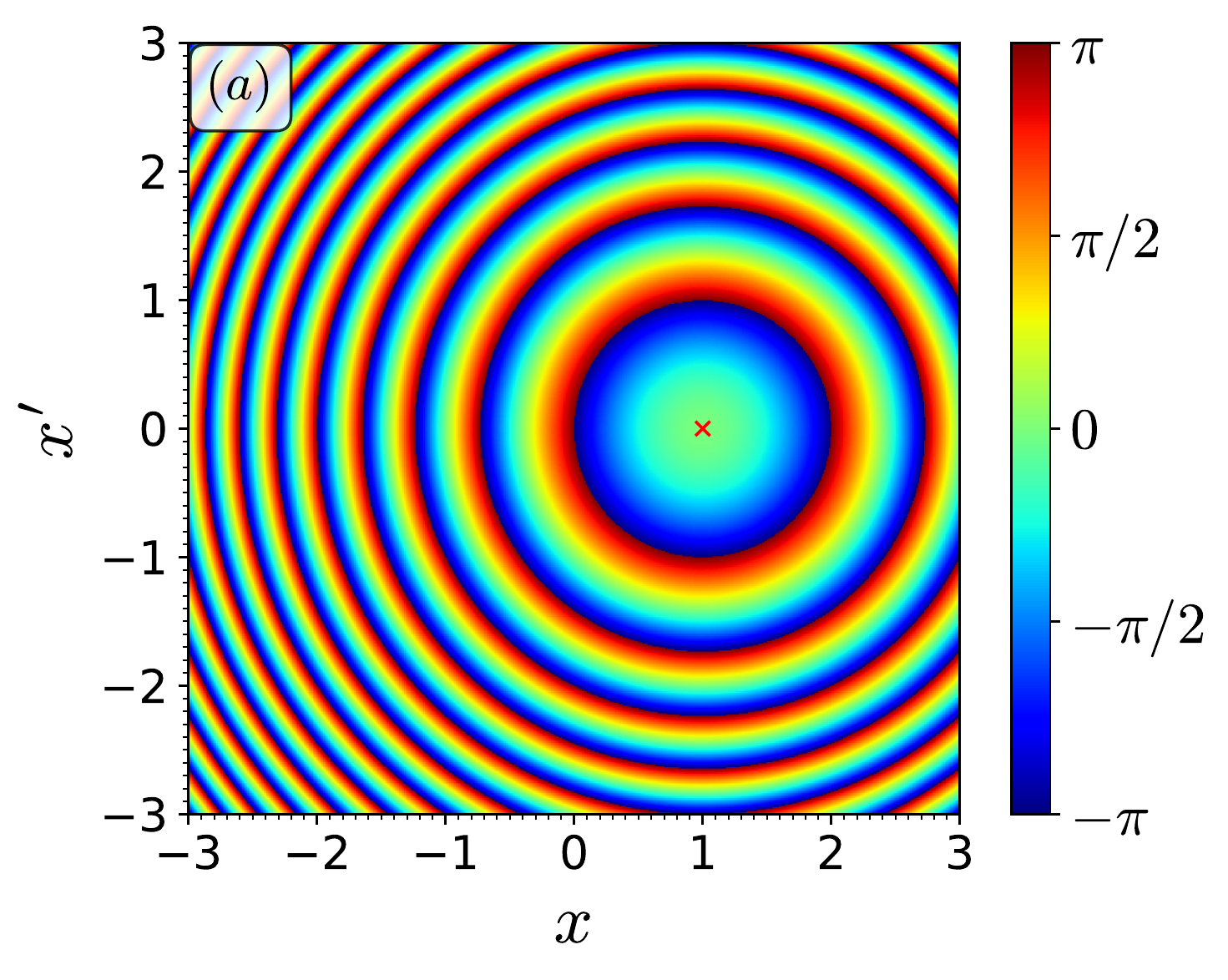}
\includegraphics[width=0.32\columnwidth]{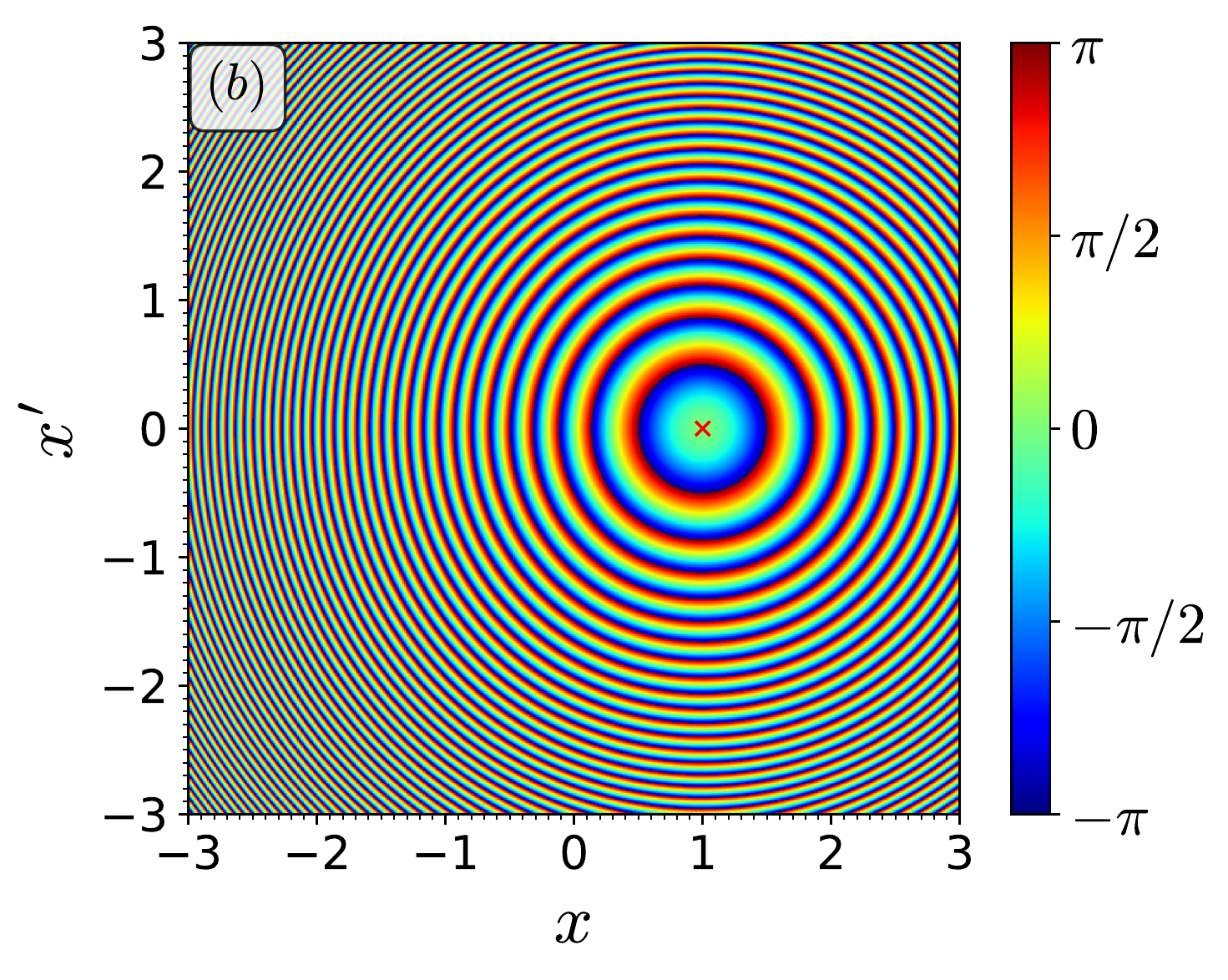}
\includegraphics[width=0.32\columnwidth]{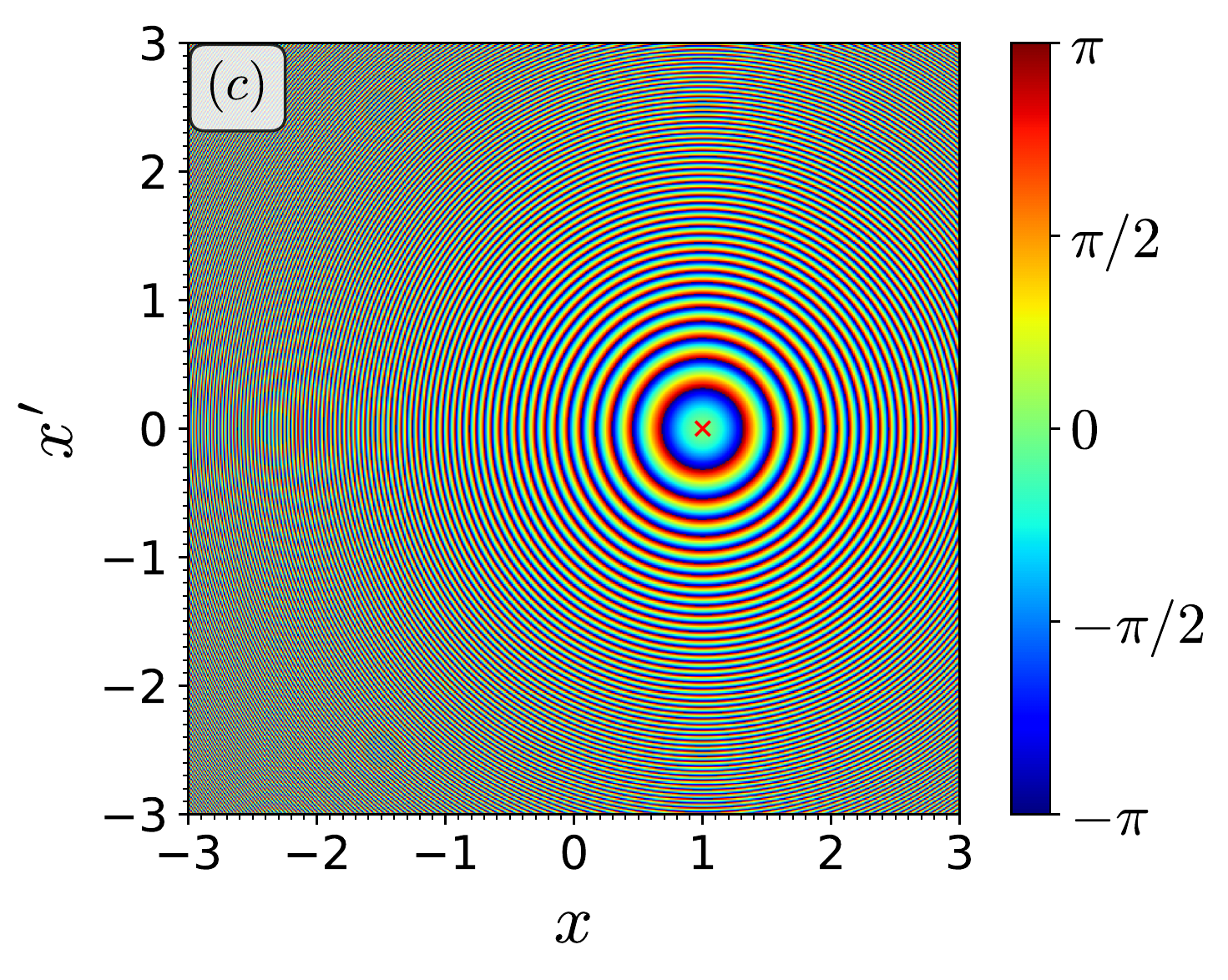}
\caption{Phase function \eqref{eq:Phi} over the lens plane for zero 
gravitational potential for different values of the Fresnel number: 
(a) $\nu=1$, (b) $\nu=4$, (c) $\nu=10$. The source located at $(1,0)$ is shown 
by a red '$\times$'. The value of its phase is chosen to be zero.
The coordinates $(x,x')$ are in units of the Einstein radius $R_{\rm E}$. }
\label{fig:phase0}
\end{figure*}

\begin{figure*}[t]
\includegraphics[width=0.32\columnwidth]{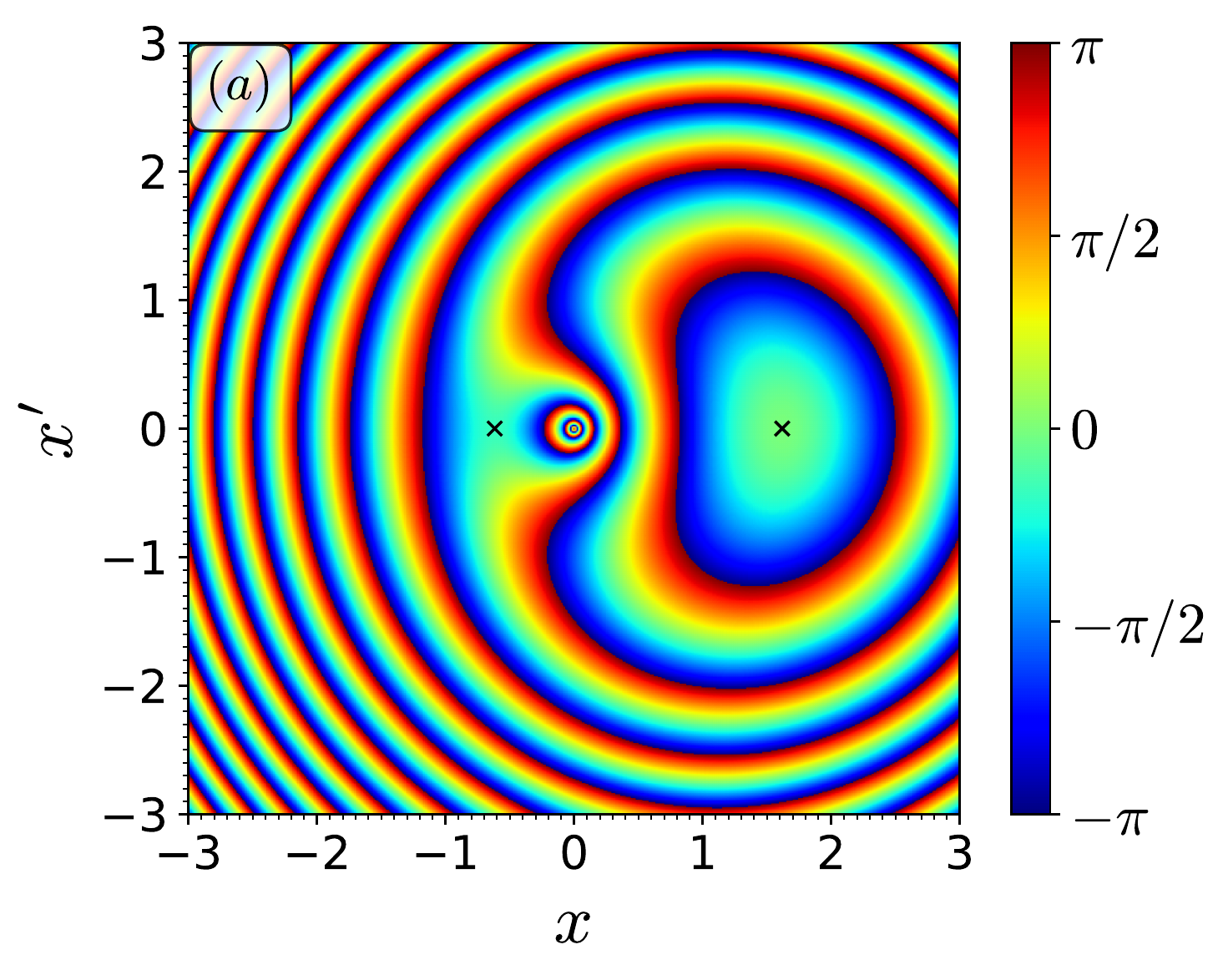}
\includegraphics[width=0.32\columnwidth]{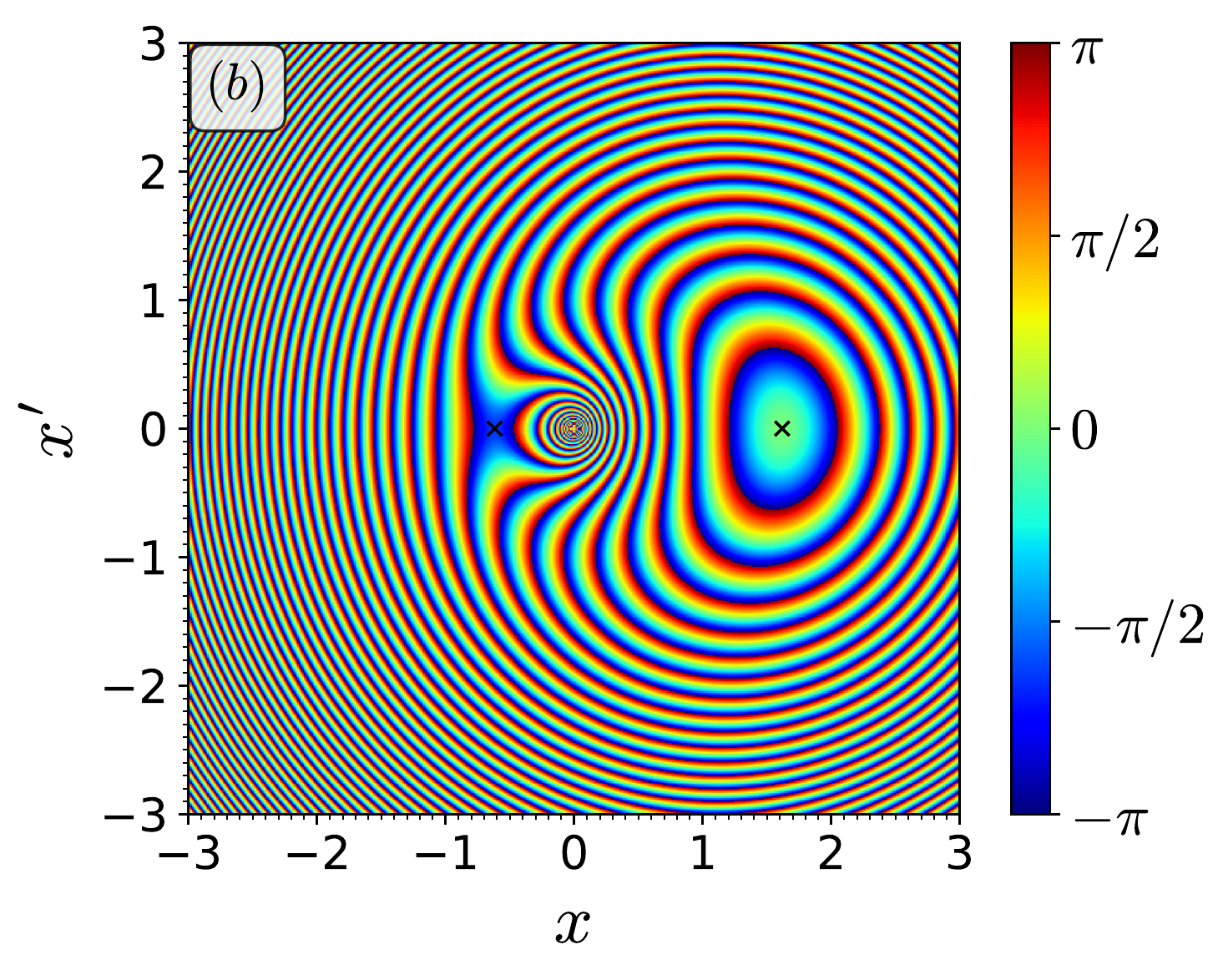}
\includegraphics[width=0.32\columnwidth]{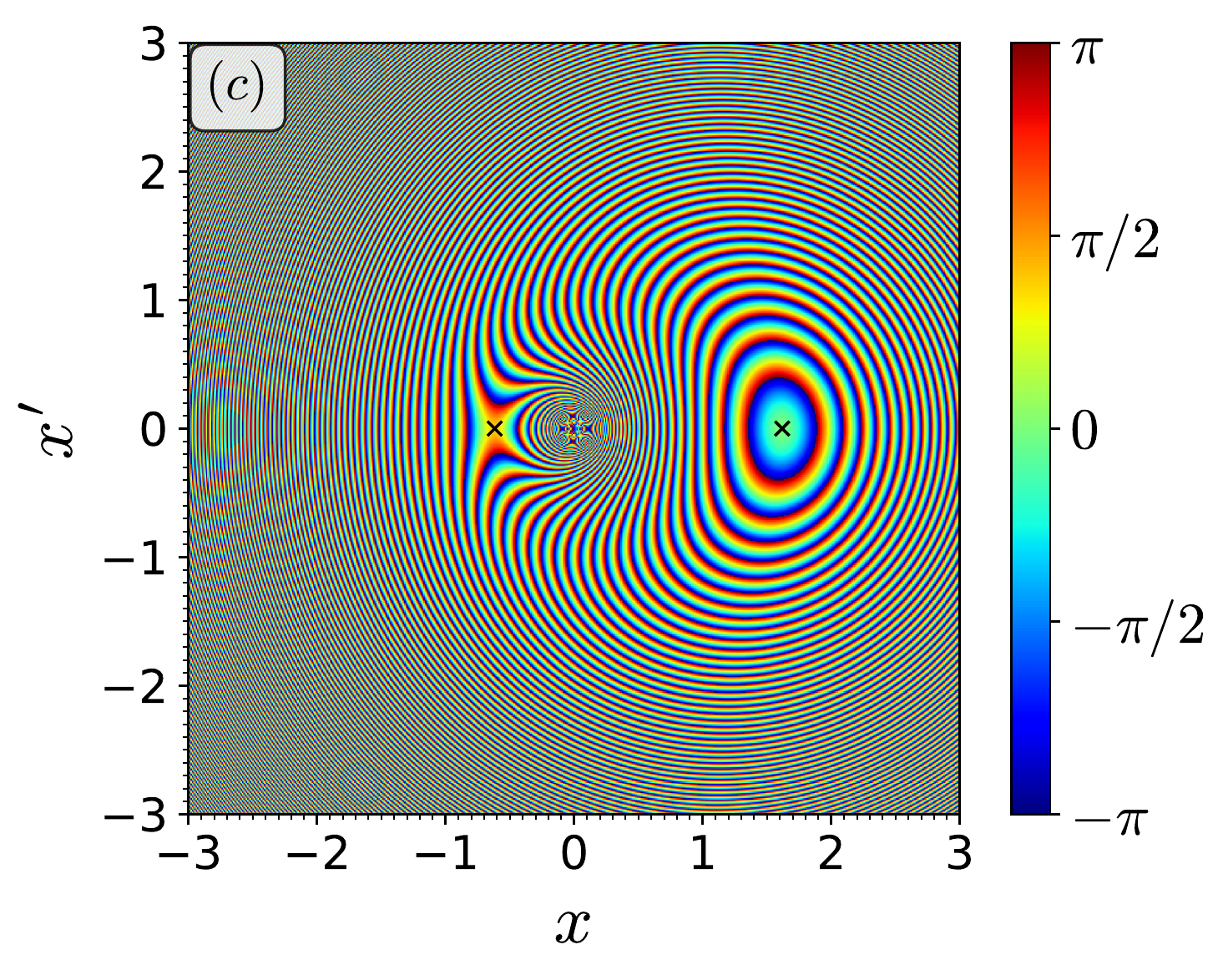}
\caption{Phase function \eqref{eq:Phi} over the lens plane for PML 
gravitational potential for different values of the Fresnel number: 
(a) $\nu=1$, (b) $\nu=4$, (c) $\nu=10$. The source located at $(1,0)$ produces 
two images (shown by black '$\times$') at $(x_1,0)$ and $(x_2,0)$ with 
$x_{1,2}=(1\pm\sqrt{5})/2$. They correspond to the minimum and the saddle 
point, respectively, of the time-delay function. The value of the phase of the first 
image is chosen to be zero. The coordinates $(x,x')$ are in units of the 
Einstein radius $R_{\rm E}$.}
\label{fig:phase}
\end{figure*}

The parameter $\nu$ introduced in Eq.~\eqref{eq:nu}
is a measure of the importance of diffraction effects in gravitational lensing 
\footnote{it is related to the parameter $w$ used in Ref.~\cite{takahashi03} by 
$w=2\pi\nu$}.
This can be seen by analyzing the partial contributions to the transmission 
factor from different parts of the lens.
As the element $d^2\mathbf{x}$ in the integral \eqref{F_nu} explores the domain 
of integration, the phase in the exponent oscillates with a rate determined by 
$\nu$.
To visualize this effect, we define a phase function $\Phi(\mathbf{x})$ on a two-dimensional domain at the lens plane
\begin{equation}
\Phi(\mathbf{x}) = \arg \left[e^{2\pi \ii \,\nu\, 
\tau(\mathbf{x},\mathbf{y})}\right], \qquad -\pi \leq \Phi \leq \pi
\label{eq:Phi}
\end{equation}
with $\mathbf{x}=(x,x')$, 
and which depends parametrically on the source position $\mathbf{y}$.
We plot this function for different values of $\nu$ by contrasting two cases: 
when the gravitational potential is turned off (Fig.~\ref{fig:phase0}) and when 
it is turned on (Fig.~\ref{fig:phase}).
If no lens is present, the lines of constant phase are circles concentric
around the (undisturbed) source at $\mathbf{x}=\mathbf{y}$. These circles, 
by design, are related to Fresnel zones \cite{born-wolf-03,jenkins-white}. 
Indeed, for a perfect alignment, $\mathbf{y}=0$, the radii $R_m$ of the 
successive Fresnel zones at the lens plane can be defined as 
\cite{jenkins-white}
\begin{equation}
m\, \frac{\lambda}{2} = R_m^2 \, \frac{d_S}{2 \,d_L d_{LS}}, \quad m=1,2,3,\dots
\end{equation}
Hence, the areas of the zones, i.e., of the rings between successive circles, are all equal to the area of the first zone, 
$\pi (R_{m+1}^2 - R_m^2) =  \pi R_1^2 \equiv \pi R_{\rm F}^2$, 
where we define
\begin{equation}
R_{\rm F} = \sqrt{\lambda \frac{d_L d_{LS}}{d_S}},
\end{equation}
which is the radius of the first Fresnel zone.
Consequently, the number of Fresnel zones covered by the circle of the Einstein radius is just the parameter 
$\nu$ introduced in Eq.~\eqref{eq:nu}
\begin{equation}
\nu = \frac{R_{\rm E}^2}{R_{\rm F}^2} =\frac{2R_{\rm S}}{\lambda},
\label{eq:NF_M}
\end{equation}
which we call the Fresnel number.
The higher $\nu$, the more phase oscillations occur inside the Einstein ring of radius $R_{\rm E}$ (the unit of length in Fig.~\ref{fig:phase0}).

For the case of lensing, when the gravitational potential is taken into account, 
the rotational symmetry is broken and multiple images appear at the lens plane 
(for the PML case, two images are seen in Fig.~\ref{fig:phase}).
It is clear that for $\nu\sim 1$, i.e., $\lambda\sim 2R_{\rm S}$, all the 
partial waves coming from the lens contribute to the transmission factor.
If however, the Fresnel number is large, $\nu\gg 1$, $|F|$ is mainly determined 
by small vicinities of the stationary points of the time delay function
(virtual images of the source), whereas the contributions from 
other parts of the lens are cancelled out by destructive interference.
In this limit, the GO approximation should be accurate 
(if $y$ is not too close to the caustic, as will be discussed below).

It is also interesting to note that the Fresnel number \eqref{eq:NF_M} for the 
PML is independent of the distance from the lens to the observer, while in case 
of lensing on a topological defect like a cosmic string, it does depend on the 
distance \cite{suyama06,pla-string16,pla-fresnel17, Jopt-mm18}.

\subsection{Full-wave solution for the transmission factor}

\begin{figure}[t]
\centering
\includegraphics[width=0.8\textwidth]{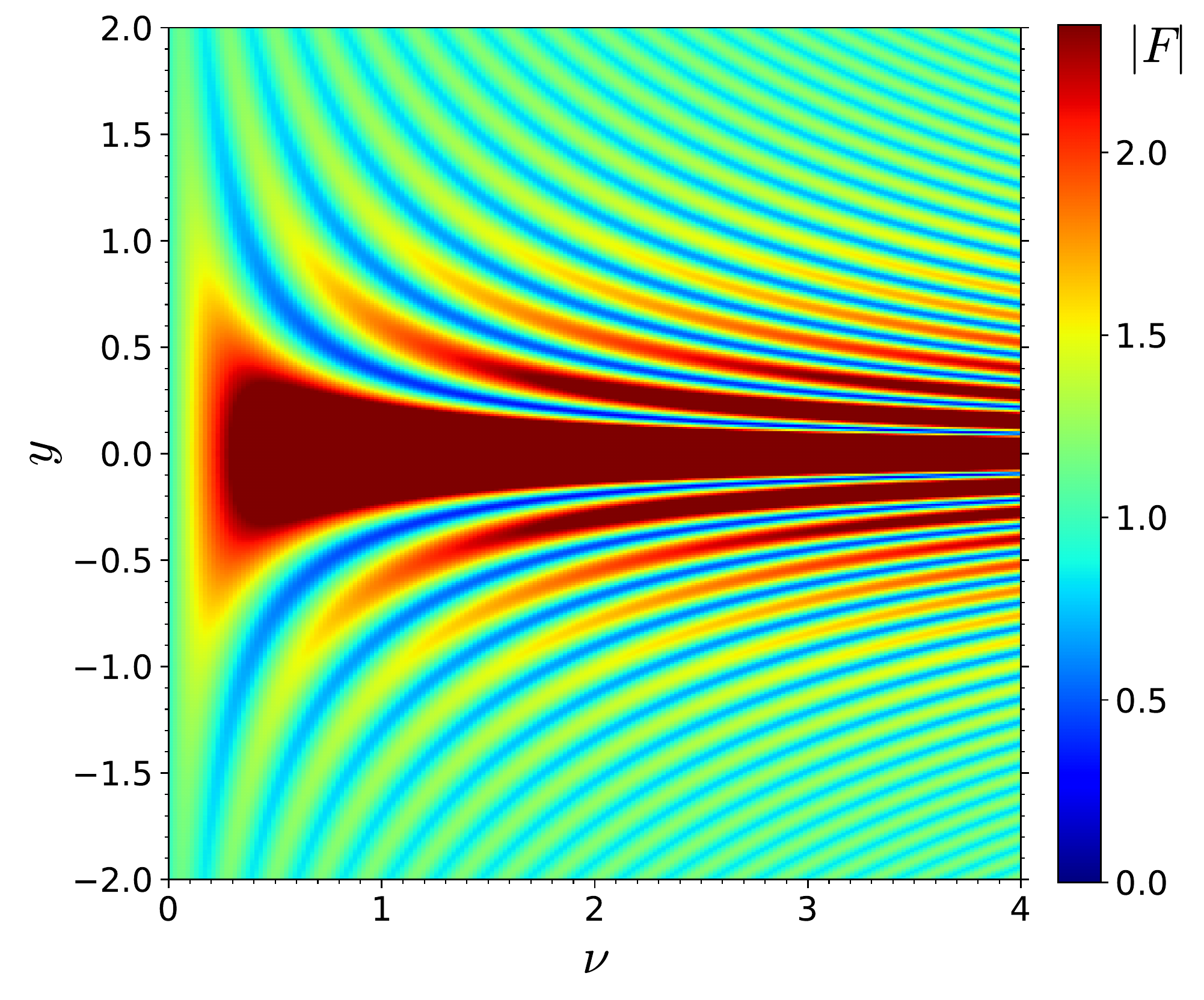}
\caption{Full-wave transmission factor $|F|$ given by Eq.~\eqref{eq:abs_F} for the PML model, as a 
function of the Fresnel number $\nu$ and source location $y$.}
\label{fig:F-nu-y}
\end{figure}

Equation \eqref{F_nu}  can be solved analytically for some simple geometries. In the particular case of PML, the transmission factor is 
obtained as \cite{schneider-92,deguchi86a}
\beqa
F = e^{\frac{1}{2}\pi s} e^{\ii s  \ln (s)} \,
\Gamma ( 1- \ii s ) \,_1F_1 ( \ii s; \,1; \,\ii s y^2 ) ,
\label{F_PML}
\eeqa
where we denoted for brevity $s\equiv \pi\nu$, $\Gamma(z)$ is the Gamma 
function, and $_1 F_1(a,b,z)$ is the confluent hypergeometric function.
It is interesting to note that the wave equation for the point mass lens 
coincides with the time-independent Schr\"odinger equation for Coulomb 
scattering \cite{deguchi86a,deguchi86b}. This means that the GWs  
will follow the same paths that charged particles would follow (at the lowest 
order) in a scattering experiment with an attracting Coulomb force. 
The exact solution for the latter was found by W.~Gordon \cite{gordon28} (see 
also \cite{mott65}) and it coincides with solution for scalar waves obtained 
from the Fresnel-Kirchhoff integral \cite{schneider-92}.
For the absolute value of the transmission factor one gets
\cite{deguchi86a}
\begin{eqnarray}
|F|=\displaystyle{
\left( \frac{2 \pi^2 \nu}{1-e^{-2 \pi^2 \nu}} \right)^{1/2}
 \left| _1F_1 ( \ii \pi \nu; \,1; \,\ii \pi \nu y^2 ) \right| },
\label{eq:abs_F}
\end{eqnarray}
which is finally a function of just two dimensionless quantities: the 
frequency (Fresnel number) $\nu$ and the source location $y$.
In Fig.~\ref{fig:F-nu-y} we depict the density plot of $|F|$ in the 
two-dimensional parameter space $(\nu,y)$. 
The figure shows clear signatures of the two-beam interference pattern.
For example, the local maxima and minima form continuous lines which 
look like hyperbolas and correspond, as will be confirmed in the subsequent 
section, to lines of constant phase between two GO rays coming from virtual 
images.

\section{Geometrical optics approximation}

\label{sec-GO}

To understand how the interference pattern in Fig.~\ref{fig:F-nu-y} is formed, 
we consider the stationary phase approximation of the Fresnel-Kirchhoff integral 
\eqref{F_nu} for which
the main contribution comes from the stationary points of the Fermat potential 
\eqref{Fermat-pot}, which are the solutions of 
\begin{equation}
\nabla_x \tau (\mathbf{x},\mathbf{y})=0.
\label{lens-eq}
\end{equation}
These solutions correspond precisely to the geometrical optics rays coming 
from the images of the source (the GO limit). The transmission factor can then 
be written as a sum over these stationary points \cite{schneider-92,nakamura99,takahashi03}: 
\begin{equation}
F_{GO}=\sum_j |\mu_j|^{1/2} e^{\ii(2\pi\nu \,\tau (\mathbf{x}_j,\mathbf{y})- n_j\pi/2)},
\label{F_GO}
\end{equation}
where 
$\mu_j=(\det \left(\partial \mathbf{y} / \partial \mathbf{x}_j 
\right))^{-1}$ is the magnification of the j-th image and $n_j=0,1,2$ is the 
Morse index for a minimum, saddle point and maximum of $\tau$, respectively.
The positions of the images $\mathbf{x}_j$ are determined by the lens equation 
\eqref{lens-eq}
\begin{equation}
\nabla_x \left[\frac{1}{2}(\mathbf{x}-\mathbf{y})^2-\psi(\mathbf{x})\right] 
=\mathbf{x}-\mathbf{y}-\nabla_x \psi(\mathbf{x})=0
\end{equation}
which is just the Fermat's principle.
For the PML model 
this equation gives $\mathbf{y}=\mathbf{x}-\mathbf{x}/|\mathbf{x}|^2$.
Without loss of generality we assume that $\mathbf{y}=(y,0)$ with $y>0$ (the 
axes can always be rotated). Thus the lens equation will have two solutions:
$\mathbf{x}_{1,2}=(x_{1,2},0)$, which correspond to two images with positions 
on the lens plane
\begin{equation}
x_{1,2}=\frac{1}{2}\,(y\pm \sqrt{y^2+4})
\label{eq:x12}
\end{equation}
and magnification
\begin{equation}
\mu_{1,2}=\displaystyle{\frac{1}{4} 
\left( \frac{y}{\sqrt{y^2+4}} + \frac{\sqrt{y^2+4}}{y} \pm 2 \right). }
\end{equation}
Taking into account that $x_1>0$ and $x_2<0$, one can define the parameter 
\begin{equation}
v \equiv \displaystyle{\frac{x_1-|x_2|}{x_1+|x_2|} = \frac{y}{\sqrt{y^2+4}}},
\end{equation}
where $0<v<1$ when $0<y<\infty$.
In terms of this parameter the magnification for each image can be written as
\begin{equation}
\mu_{1,2}=\displaystyle{\frac{1}{4} \left( v + \frac{1}{v} \pm 2 \right)
= \frac{1}{4}\left( \sqrt{v} \pm \frac{1}{\sqrt{v}} \right)^2}
\end{equation}
and it is seen that $\mu_1$ and $\mu_2$ are both positive.
For the PML model which has two images (minimum and saddle point),  
the transmission factor \eqref{F_GO} becomes
\begin{equation}
F_{GO}= \sqrt{\mu_1}\,e^{\ii \phi_1} + \sqrt{\mu_2} \,e^{\ii\phi_2-\ii\phi_M},
\end{equation}
where $\phi_i = 2\pi \nu \,\tau(x_i,y)$ is the phase of each image and 
$\phi_M=\pi/2$ is the Morse (topological) phase shift of the second 
image (saddle point). Thus, we get
\begin{equation}
F_{GO}= \left(\sqrt{\mu_1} + \sqrt{\mu_2} \,e^{2 \ii \alpha} \right) e^{\ii \phi_1},
\label{F_GO_PML}
\end{equation}
with 
\begin{equation}
\alpha = \pi \nu \,\tau_{21} - \pi/4,
\label{eq:alpha}
\end{equation}
which depends on the time delay between the two images 
$\tau_{21}\equiv\tau(x_2,y)-\tau(x_1,y)$. For the latter we obtain
\begin{equation}
\tau_{21} = \displaystyle{ \frac{2v}{1-v^2} + \ln  
\left( \frac{1+v}{1-v} \right)} 
\equiv \tau_{geom} + \tau_{grav}.
\label{tau_21}
\end{equation}
It takes into account two effects: the geometrical time delay (the first 
term) and the gravitational one, or Shapiro delay (the second term). 
It is clear that $\tau_{21}>0$ since $x_1$ corresponds to the minimum travel 
time.
The absolute value (squared) of the transmission factor \eqref{F_GO_PML} can 
finally be written as
\begin{align}
|F_{GO}|^2 
&=& \mu_1 + \mu_2 + 2 \sqrt{\mu_1 \mu_2} \,\cos \,2\alpha \nonumber \\ 
&=&  v\, \sin^2 \alpha + (1/v) \,\cos^2 \alpha \nonumber \\
&=&  \frac{ y^2 +4 \cos^2 \alpha}{y\sqrt{y^2+4}}.
\label{F_GO_sin}
\end{align}
It should give an interference pattern for which the fringe spacing is determined by $\alpha$, 
while the minimum and maximum values of $|F_{GO}|^2$ are equal to $v$ and $1/v$, respectively.
Another important observation is that the beating oscillations can be 
significantly amplified and their amplitude is more pronounced for small values 
of $y$ when the source approaches the caustic. 
Strictly at $y=0$, Eq.~\eqref{F_GO_sin} is not valid since it diverges. 
The smallest value of $y$ for which the GO approximation is still accurate 
will be obtained from the full-wave solution in the next section and it depends on $\nu$. 

\begin{figure}[t]
\centering
\includegraphics[width=0.7\columnwidth]{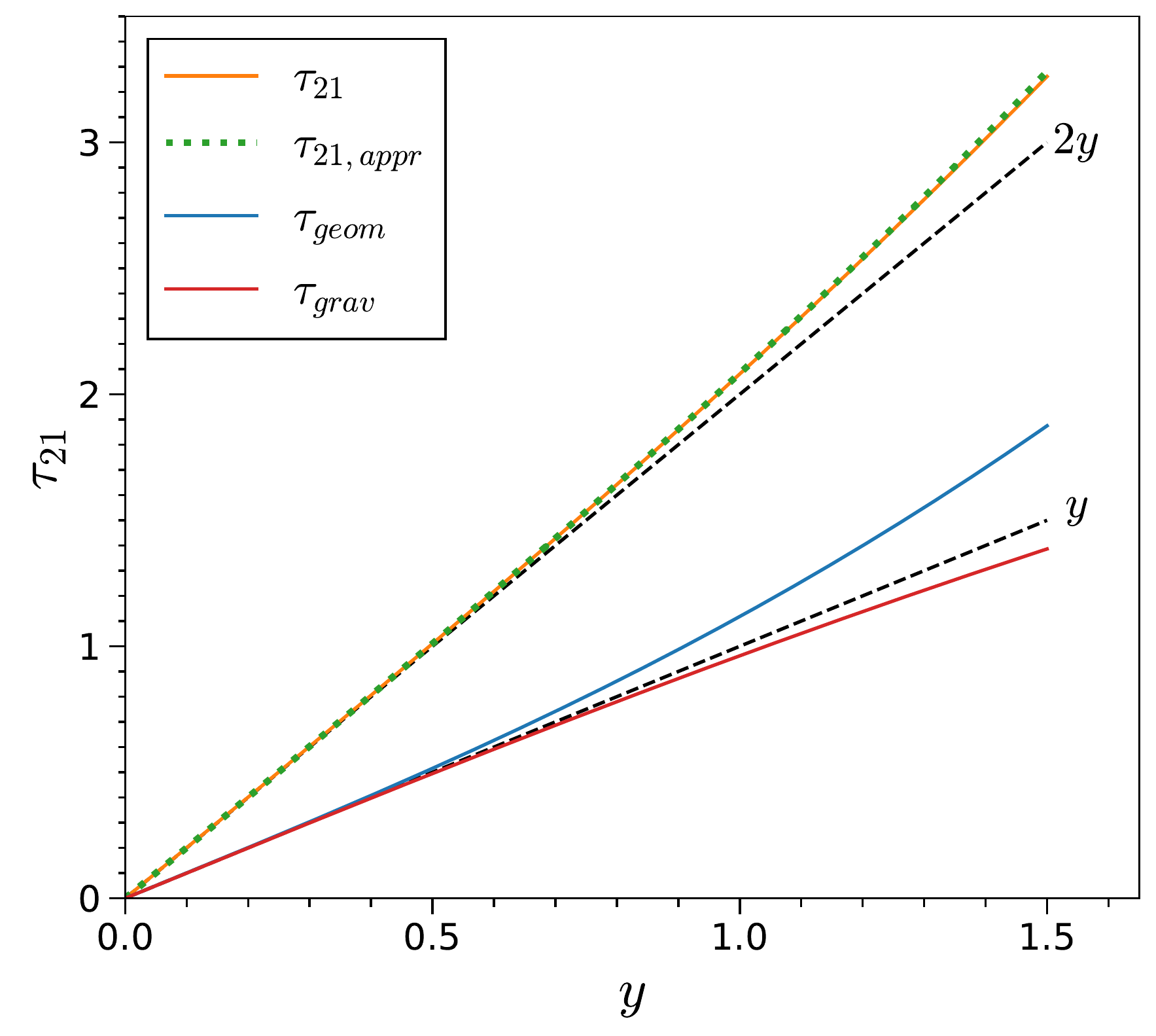}
\caption{Time delay $\tau_{21}$ given by Eq.~\eqref{tau_21} as a function 
of the source location $y$ (orange line). It is composed of geometrical (blue 
line) and gravitational (red line) constituents.
The asymptotic approximations are shown for comparison: at first order 
$\tau_{21} \approx 2y$, 
and at the next order $\tau_{21} \approx 2y+y^3/12$ (green dotted line).}
\label{fig:tau21}
\end{figure}

For small $y$ we can expand \eqref{tau_21} in Taylor series to obtain:
\begin{equation}
\tau_{21}(y) = 2y + \frac{1}{12}y^3 + O(y^5).
\label{tau_Taylor}
\end{equation}
In Fig.~\ref{fig:tau21} we compare the time delay \eqref{tau_21} with its 
asymptotic approximation \eqref{tau_Taylor} and we also present the 
geometrical and gravitational partial contributions for completeness.
It is clearly seen that for $0<y\lesssim 0.5$ (close alignment) one can safely take only the 
leading-order term, $\tau_{21}\approx 2y$, whereas for $0.5\lesssim y \lesssim 
1.5$ the next-order term is needed, $\tau_{21}\approx 2y+y^3/12$, to reproduce the
exact value with sufficiently high accuracy. These approximations substantially 
simplify the analytical treatment of the transmission factor in subsequent 
chapters.
The rather wide range of validity of $\tau_{21}\approx 2y$ 
can be explained by the fact that the cubic terms coming from the geometrical 
and gravitational delays mutually almost compensate each other being of the 
opposite sign.
The factor of $2$ is a consequence of equal contributions from geometrical and 
gravitational parts to the linear term.

\section{Full-wave vs geometrical optics}
\label{Sec_full-GO}
\begin{figure}[t]
\centering
\includegraphics[width=0.75\columnwidth]{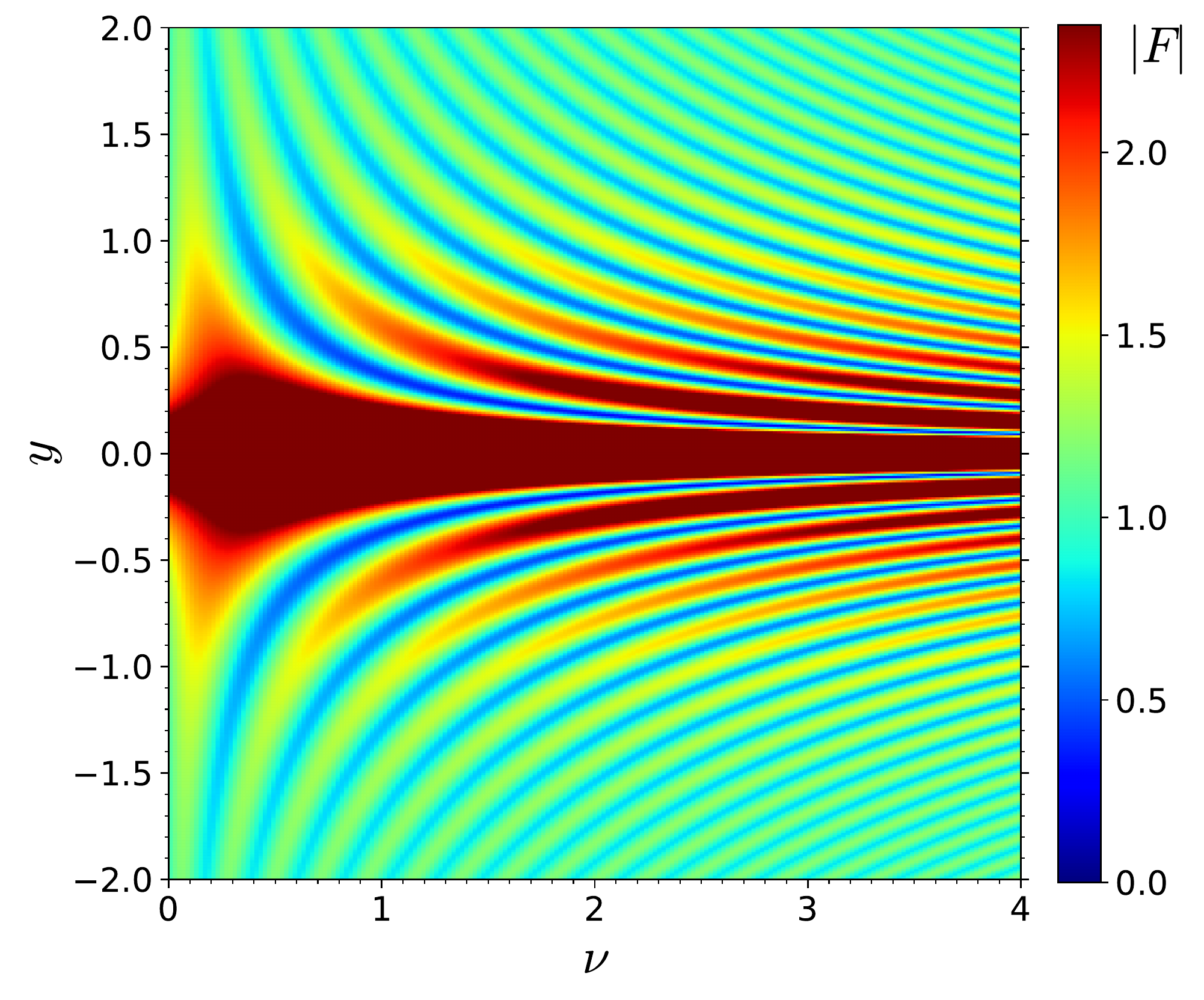}
\includegraphics[width=0.75\columnwidth]{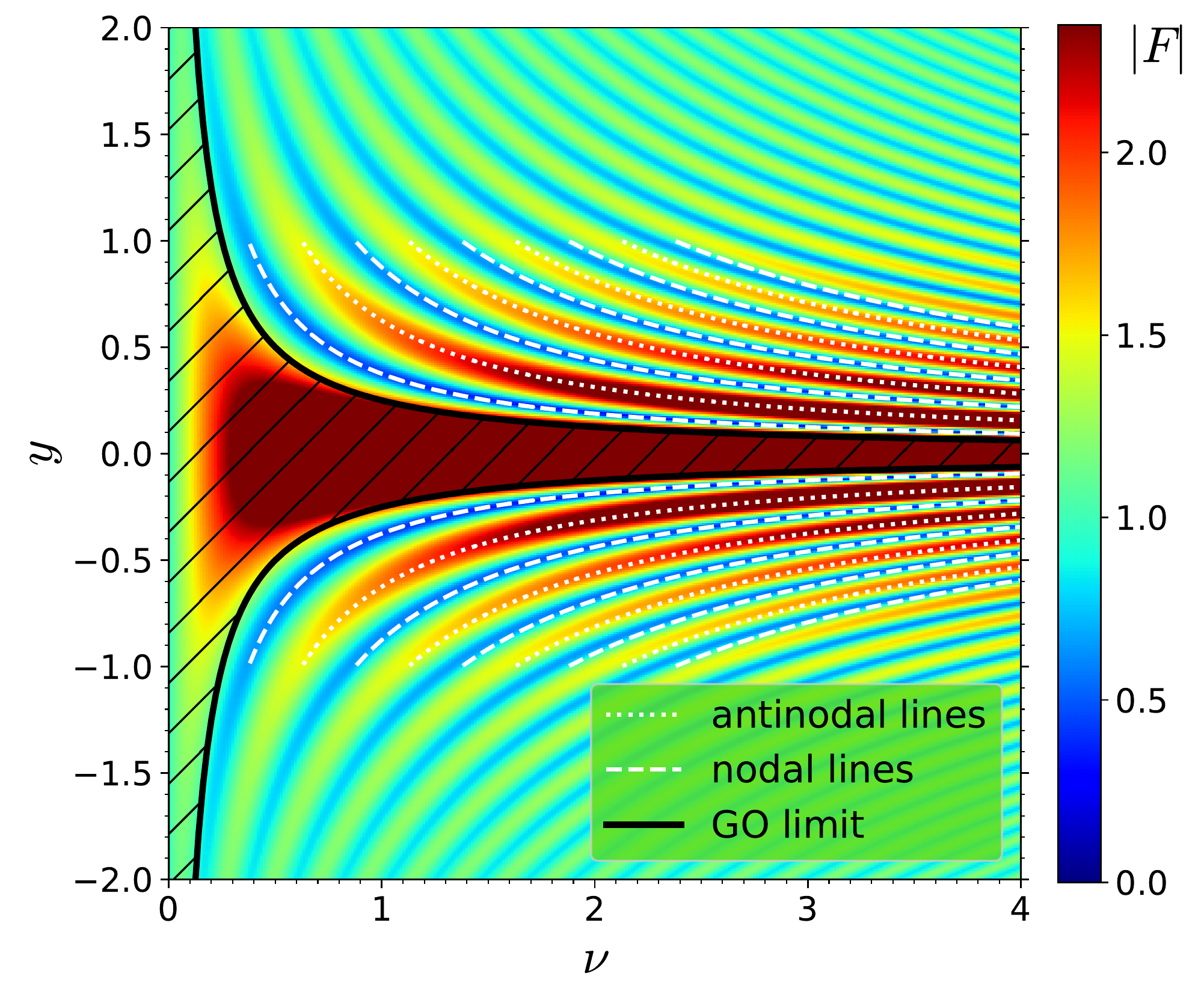}
\caption{
Wave pattern for the transmission factor $|F|$ calculated for the PML as a 
function of frequency $\nu$ and source location $y$:
(a) geometrical-optics (GO) approximation; 
(b) full-wave solution with superimposed lines of constant phase between GO 
rays (white dotted and dashed lines).
The excluded masked region where the GO approximation is not valid is bounded 
by a black line.}
\label{fig:F-compare}
\end{figure}

Next we compare in Fig.~\ref{fig:F-compare} the GO transmission factor 
\eqref{F_GO_sin} with the full-wave solution \eqref{eq:abs_F}. 
It is seen that the interference fringes are well reproduced by GO rays in the 
whole parameter space except at small values of $\nu$.
We also expect the discrepancy between the figures (not seen in the density 
plots) in the region close to the caustic $y= 0$ where the GO approximation diverges. 

\subsection{Nodal and antinodal lines}

The lines of local maxima and minima in the interference pattern 
can be easily understood if we regard them as constant-phase lines (nodal and 
antinodal lines \cite{pla-fresnel17,Jopt-mm18}) which appear due to interference 
between two GO rays.
Indeed, to have constructive or destructive interference, 
the maxima and minima occur when the ``optical path difference'' 
$c\Delta t_{21}$ between the GO rays can be written in terms of wavelength 
$\lambda$ as follows
\begin{equation}
c\Delta t_{21} = 
\begin{cases}
n \lambda + \frac{\lambda}{4}, \quad \text{at maxima,} \\ 
n \lambda + \frac{\lambda}{2}+\frac{\lambda}{4},  \quad \text{at minima}
\end{cases}
\label{eq:maxmin}
\end{equation}
with $n=0,1,2,\dots$. Here, the additional term $\lambda/4$ takes into account 
the Morse phase shift $\pi/2$ between two stationary points---the minimum 
and the saddle point---as follows from Eq.~\eqref{F_GO} (see also 
Fig.~\ref{fig:phase}). 
Taking into account that $c\Delta t_{21}= 2 R_{\rm S} \tau_{21} = \lambda \nu \,\tau_{21}$, Eq.~\eqref{eq:maxmin} in terms of dimensionless parameters reads
\begin{equation}
\nu \tau_{21}(y) = \displaystyle{
\begin{cases}
\left( n+\frac{1}{4} \right), \quad \text{at maxima,} \\[1mm]
\left( n+\frac{3}{4} \right),  \quad \text{at minima}.
\end{cases}
}
\label{eq:hyperbolas_general}
\end{equation}
For the close alignment condition, $y\lesssim 0.5$, substituting $\tau_{21}\approx 2y$, we obtain
\begin{equation}
y_n = \displaystyle{\frac{1}{2\nu}\cdot
\begin{cases}
\left( n+\frac{1}{4} \right), \quad \text{at maxima,} \\[1mm]
\left( n+\frac{3}{4} \right),  \quad \text{at minima},
\end{cases}
}
\label{eq:hyperbolas}
\end{equation}
which are hyperbolas in $(\nu,y)$ parameter space.
The same formulas can formally be obtained by assuming that the 
phase $\alpha$ in Eq.~\eqref{F_GO_sin} satisfies
\begin{equation}
\alpha = 
\begin{cases}
\pi n, \quad \text{at maxima,} \\ 
\frac{\pi}{2} + \pi n,  \quad \text{at minima}.
\end{cases}
\end{equation}
The hyperbolas \eqref{eq:hyperbolas} are shown in Fig.~\ref{fig:F-compare}(b) 
superimposed on the wave pattern (shown in color) obtained from full-wave 
solution \eqref{eq:abs_F} (the same as in Fig.~\ref{fig:F-nu-y}).
A good agreement is observed for a rather wide range of the source location 
\footnote{we only omit the very first line of maximum for $n=0$ which is outside 
the region of the GO validity.}
that validates the approximation $\tau_{21}\approx 2y$ for close alignment at $0<y\lesssim 0.5$.

\begin{figure}[t!]
\centering
\includegraphics[width=0.7\columnwidth]{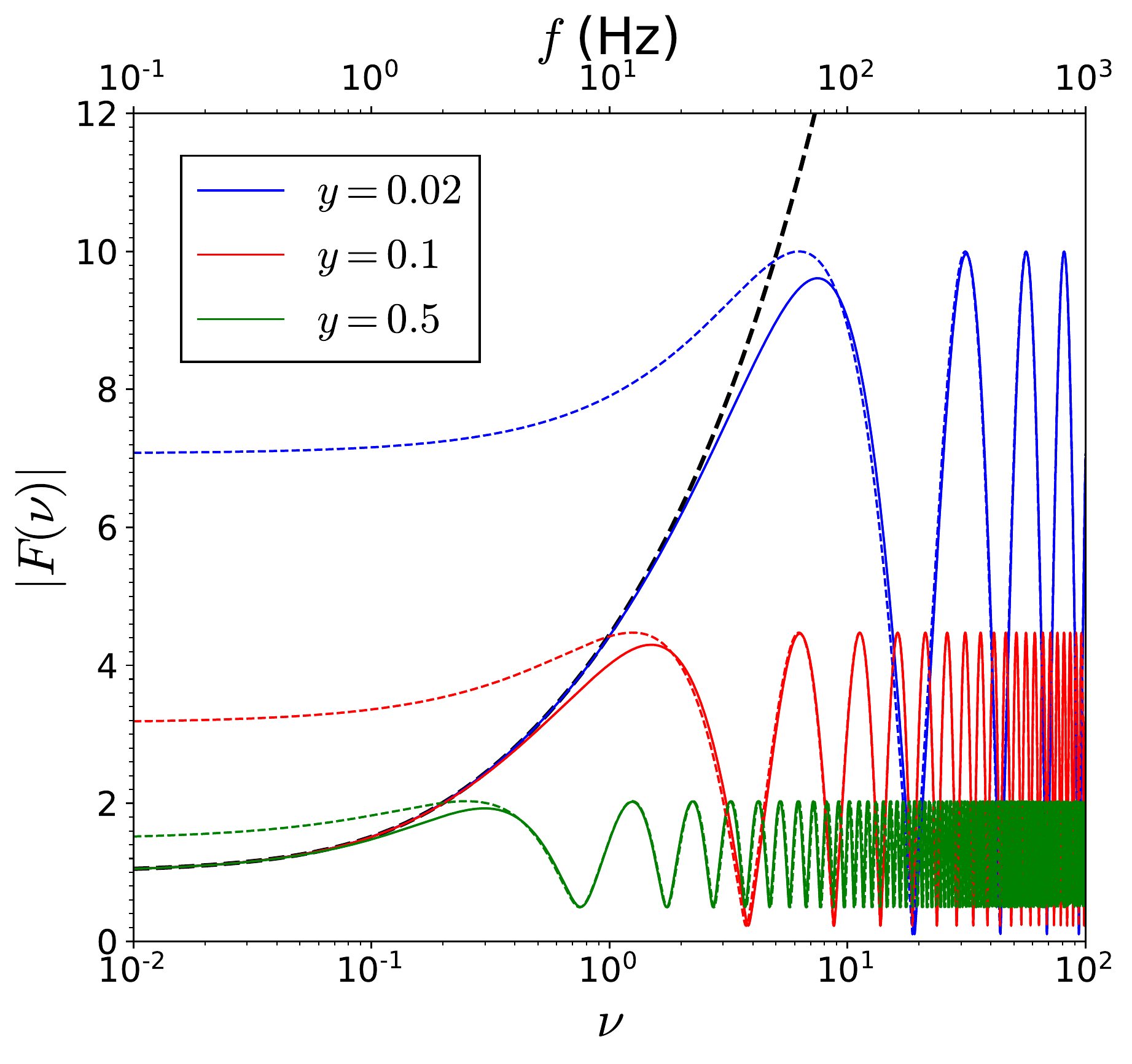}
\caption{Transmission factor $|F(\nu)|$ vs frequency for different angular 
locations of the source:
$y=\theta_S/\theta_{\rm E}=0.02; \;0.1;\; 0.5$. 
Solid lines correspond to the full-wave solution \eqref{eq:abs_F} and 
dashed lines, represented by the same colour, to the GO approximation 
\eqref{F_GO_sin}. The asymptotic low-frequency curve 
\eqref{asymptotic_lowfreq} is shown by black dashes.
The bottom horizontal axis shows the dimensionless frequency (or Fresnel 
number) $\nu$ defined in Eqs.~\eqref{eq:nu} and \eqref{eq:NF_M}, 
and the top axis maps into a physical GW frequency $f$ (in Hz) in the LIGO/Virgo 
band for a specific choice of lens mass $M=5\times 10^3 M_\odot$.}
\label{fig:F-nu}
\end{figure}

\subsection{Transmission factor vs frequency}
\label{sec-F-nu}


To get more insight into how the transition from full-wave to GO approximation 
occurs, we present in Fig.~\ref{fig:F-nu} 
plots of $|F(\nu)|$ 
for some fixed values of the source position $y$ (horizontal sections 
of Fig.~\ref{fig:F-compare}).
This would correspond to a situation when the relative displacement of the lens 
and the source is negligible during the observational time, which for 
ground-based GW detectors is $\lesssim$ seconds, while for space-based detectors 
and pulsar timing arrays can reach larger timescales, $\lesssim$ years. 
It is seen from Fig.~\ref{fig:F-nu} 
that at low frequencies (small Fresnel numbers $\nu$), the GO approximation substantially overestimates the transmission factor, which means that other regions of the lens apart from the GO images (stationary points) 
contribute significantly 
and the stationary phase approximation does not hold.
As a result the amplification is suppressed.
In this limit, full-wave solutions should be used which are seen to converge to 
a unique asymptotic curve
\begin{equation}
|F(\nu)|_{\rm max} = \displaystyle{
\left( \frac{2 \pi^2 \nu}{1-e^{-2 \pi^2 \nu}} \right)^{1/2}},
\label{asymptotic_lowfreq}
\end{equation}
which is obtained from \eqref{eq:abs_F} in the limit $\nu y \to 0$.
Note that Eq.~\eqref{asymptotic_lowfreq} represents also the transmission 
factor for the particular case $y = 0$, when the source, lens and observer are 
perfectly aligned. 
It gives the maximum amplification that the GWs may reach for each frequency, which is achieved at the caustic.

\begin{figure}[t]
\centering
\includegraphics[width=0.7\columnwidth]{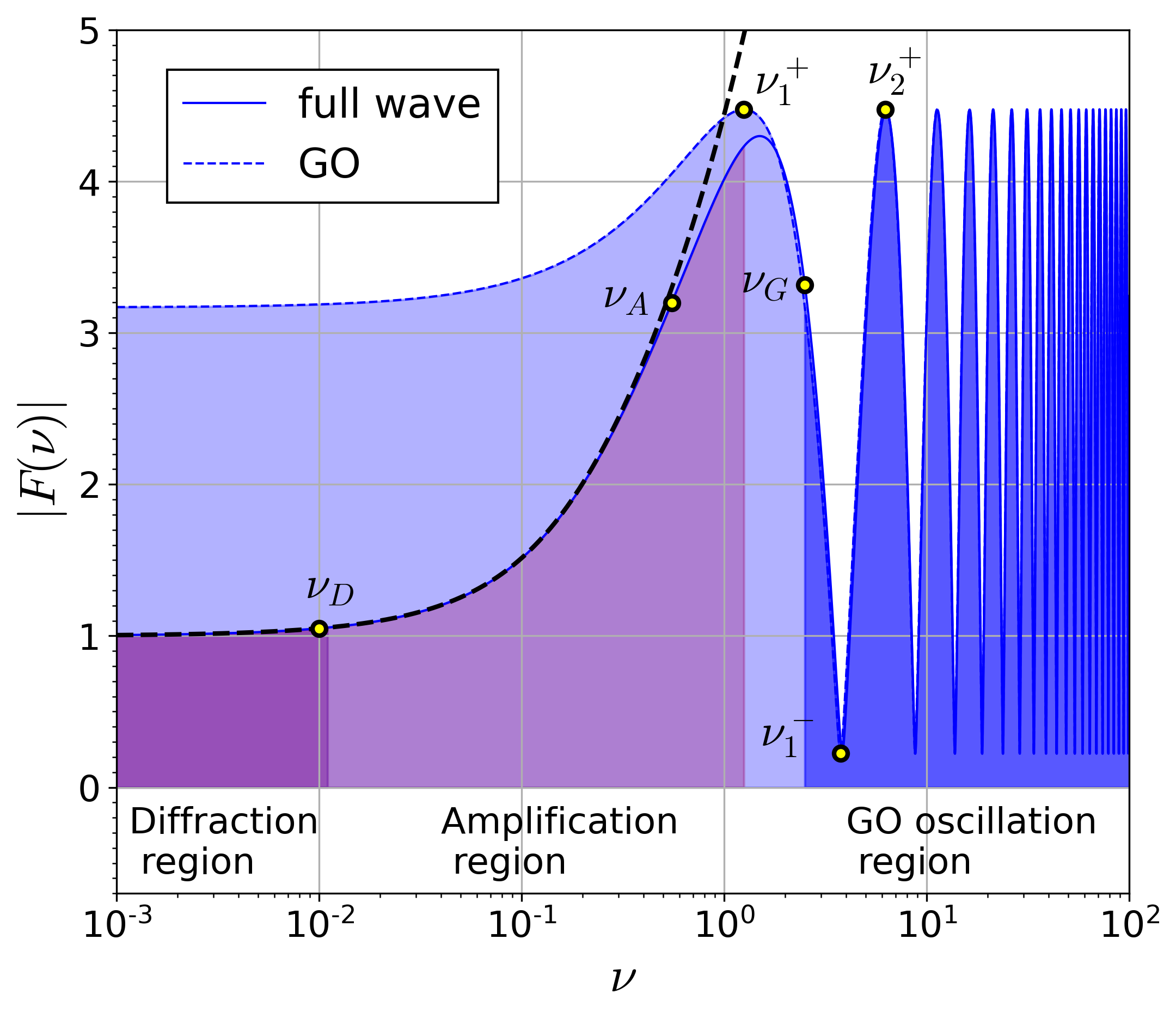}
\caption{Transmission factor $|F(\nu)|$ vs dimensionless frequency (Fresnel number) $\nu$ for the source location at $y=0.1$.
The full-wave and the GO solutions are compared.
The asymptotic low-frequency curve \eqref{asymptotic_lowfreq} is shown by black dashes. 
Characteristic frequencies and regions are indicated.}
\label{fig:F-nu-1}
\end{figure}
In the opposite limit of high frequencies, the full-wave solutions 
asymptotically coalesce into the corresponding GO curves and one can observe 
quite similar oscillation patterns for different $y$.
To describe in more detail the transition from wave optics to GO regime, we 
define some characteristic frequencies in Fig.~\ref{fig:F-nu-1} where $|F(\nu)|$ 
is depicted for a particular value of $y=0.1$ (close alignment). 
We distinguish the following regions of interest: 
\\[2mm]\mbox{\bf (i) Diffraction region, $\nu \lesssim \nu_D$.}\\[2mm]
In this region the wavelength $\lambda$ is too large with respect to the 
Schwarzschild radius, so the lens does not affect the propagation of waves,
$|F|\approx 1$ for any $y$. 
To estimate the upper limit of this region, we expand the function 
\eqref{asymptotic_lowfreq} at small arguments $|F(\nu)|_{\rm max} = 
1+\frac{1}{2}\pi^2\, \nu + O(\nu^2)$, to obtain $\nu_D\ll 2/\pi^2$, so $\nu_D 
\simeq 0.01$ is a reasonable value.
\\[2mm]\mbox{\bf (ii) Amplification region, $\nu_D \lesssim \nu<\nu_1^+$.}\\[2mm]
This region extends from $\nu_D$ up to the first maximum and it is 
characterized by a monotonic magnification without oscillations as $\nu$ 
increases. 
Although the full-wave and GO solutions still have a discrepancy at the first 
maximum, the values of  $\nu$ are close, so for simplicity we can take as the 
upper limit of this region the first GO maximum given by 
Eq.~\eqref{eq:hyperbolas} for $n=0$ at $\nu_1^+ = 1/(8y)$. 
One can also use for the first full-wave maximum an adjustment formula given in Ref.~\cite{urrutia21}.
As $\nu$ increases, the full-wave solution follows closely the asymptotic curve 
\eqref{asymptotic_lowfreq} in the interval: $\nu_D \lesssim \nu \lesssim 
\nu_A$, where the point of separation $\nu_A$ can be estimated by applying to 
Eq.~\eqref{eq:abs_F} the asymptotic expansion
$\,_1F_1 (\ii\pi\nu;\,1; \,\ii\pi\nu y^2) \approx J_0(2\pi\nu y) = 1 - 
\pi^2\nu^2 y^2 + O(\nu^4 y^4)$, where the Bessel function $J_0$ is expanded at 
small arguments. 
The full-wave solution starts to deviate from $|F(\nu)|_{\rm max}$
when the second term in the expansion is non-negligible.
This happens approximately at $\pi^2\nu_A^2 \, y^2 \simeq 0.03$, so $\nu_A \simeq 
\sqrt{0.03}/(\pi y) \simeq 0.055/y$.
It is seen that the value of $\nu_A$ increases as $y$ decreases, 
in accordance with Fig.~\ref{fig:F-nu}.
\\[2mm]\mbox{\bf (iii) GO oscillation region,  $\nu>\nu_G$.}\\[2mm]
We define $\nu_G$ as a threshold frequency at which the full-wave and GO 
solutions start practically to coincide. As seen from the figure, it should be 
in between the first maximum $\nu_1^+$
and the first minimum $\nu_1^-$,
so we define it for simplicity in the middle, $c\Delta t_{21}\lvert_{\nu=\nu_G}=\lambda/2$, that corresponds to
\begin{equation}
\nu_G = \frac{1}{2 \tau_{21}(y)} \approx \frac{1}{4y},
\label{eq:nuG}
\end{equation}
where the approximation is given under the close alignment $y\lesssim 0.5$. 
The smaller the value of $y$, the higher the frequency $\nu_G$. In the limit $y\to 0$, one gets 
$\nu_G\to\infty$, this means that for the line of sight the GO approximation is never 
valid, since $y=0$ corresponds to the caustic point.
The region of the GO validity in the two-parameter space $(\nu,y)$
is indicated in Fig.~\ref{fig:F-compare}(b).
For higher values of $y$, the next-order terms in the expansion of $\tau_{21}$ are needed to obtain $\nu_G$.
In particular, for $0.5\lesssim y \lesssim 1.5$, 
$\tau_{21}\approx 2y+y^3/12$ can be used to obtain the onset, otherwise one can use the general formula \eqref{tau_21}.
It should be noted that the approximations for $\tau_{21}$ substantially reduce the computational time when one needs to match the full-wave and GO solutions at some value in the data analysis techniques. 
The authors of Ref.~\cite{urrutia21} suggested for the onset of the GO an adjustment formula for the first maximum, but, as seen from Figs.~\ref{fig:F-nu} and \ref{fig:F-nu-1}, the two solutions do not coincide well there, so the first maximum may be used for rough estimations, but not as a matching point. Additionally, we note that instead of the middle point \eqref{eq:nuG}, the second or the third maximum can also be used as a definition of the onset: the higher the frequency, the better the GO solution matches the full-wave one. However, Eq.~\eqref{eq:nuG} gives the lower bound for the matching point.

We can further simplify our analytical treatment by taking into account that 
the range $\nu> \nu_G$ with $y\lesssim 0.5$ is of the most interest, since 
for these values two images are well pronounced, they are of comparable 
intensity and the interference pattern is strongly amplified.
In this case, by expanding the GO solution \eqref{F_GO_sin} we obtain a reduced 
analytical formula
\begin{equation}
|F_{GO}|^2 \approx \frac{y}{2} + \left(\frac{2}{y}-\frac{y}{4}\right) \, \cos^2 
\left[2\pi\nu \,y -\frac{\pi}{4}\right],
\label{cos_approx}
\end{equation}
which reproduces quite well the oscillation pattern.
Note that for any fixed $y\neq 0$, the function $|F(\nu)|$ oscillates 
with evenly spaced maxima and minima. The values of minima are equal to $y/2$ 
(to the leading order). They are not zero because the interference is not 
completely destructive when the source is off the line of sight and the images 
are therefore of different magnifications.
The amplitude of the oscillations (the difference between maximum and minimum 
values) increases $\sim 2/y$ when $y\to 0$ (see Fig.~\ref{fig:F-nu}). 
These oscillations appear due to crossing the nodal and antinodal lines---the 
constant-phase lines between the GO rays---when 
$\nu$ varies (Fig.~\ref{fig:F-compare}). 
As will be shown below, they also determine the interference fringe 
in the GW lensed waveform.
The fringe spacing over the spectrum is uniform with the characteristic interval
\begin{equation}
\Delta \nu = \frac{1}{\tau_{21}(y)} \approx \frac{1}{2y},
\label{fringe}
\end{equation}
i.e., to the leading order, it is inversely proportional to the parameter $y$.
Translating into physical units, the frequency spacing of the fringe is 
determined by
\begin{equation}
\Delta f = \frac{1}{2\, t_M\,y} 
\simeq 2.5\times 10^4 \,{\rm Hz} \,\left(\frac{M_\odot}{M} \right)
\,\left(\frac{\theta_{\rm E}}{\theta_S} \right).
\label{fringe_units}
\end{equation}
with
\begin{equation}
f_n = \displaystyle{\frac{1}{2\, t_M\,y}\cdot
\begin{cases}
\left( n+\frac{1}{4} \right), \quad \text{at maxima,} \\[1mm]
\left( n+\frac{3}{4} \right),  \quad \text{at minima}.
\end{cases} }
\label{eq:hyperbolas_physical}
\end{equation}
For $y = 0.5$ and $M=100 M_\odot$ this gives $\Delta f = 500\,$Hz.
Note that the fringe spacing for close alignment, $y\lesssim0.5$ depends on the product 
 $2\,t_M\,y$. For higher values of $y$, the product 
would be $t_M\,\tau_{21}$, which corresponds to the time delay between the two images, $\Delta t_{21}$. 


\subsection{Transmission factor vs source location $y$}
\label{sec-F-y}

Another situation of interest is the case of a continuous monochromatic signal 
coming for instance from (i) an isolated neutron star emitting GWs at 
frequencies within the LIGO/Virgo band, or (ii) a supermassive black hole (SMBH) binary 
in their long inspiral phase, so that their frequencies can be considered to be 
almost constant for a long time \cite{maggiore-18}.
The GWs emitted by the latter sources fall in the frequency band of the space 
interferometer LISA and pulsar timing arrays (PTA) experiments.
In this situation, we assume that $\nu$ is approximately constant, but the angular 
parameter $y$ changes with time due to the relative displacement of the source, 
lens, or observer \cite{oguri19,liao19,sugiyama20,jow20}.

\begin{figure}[t]
\centering
\includegraphics[width=0.75\columnwidth]{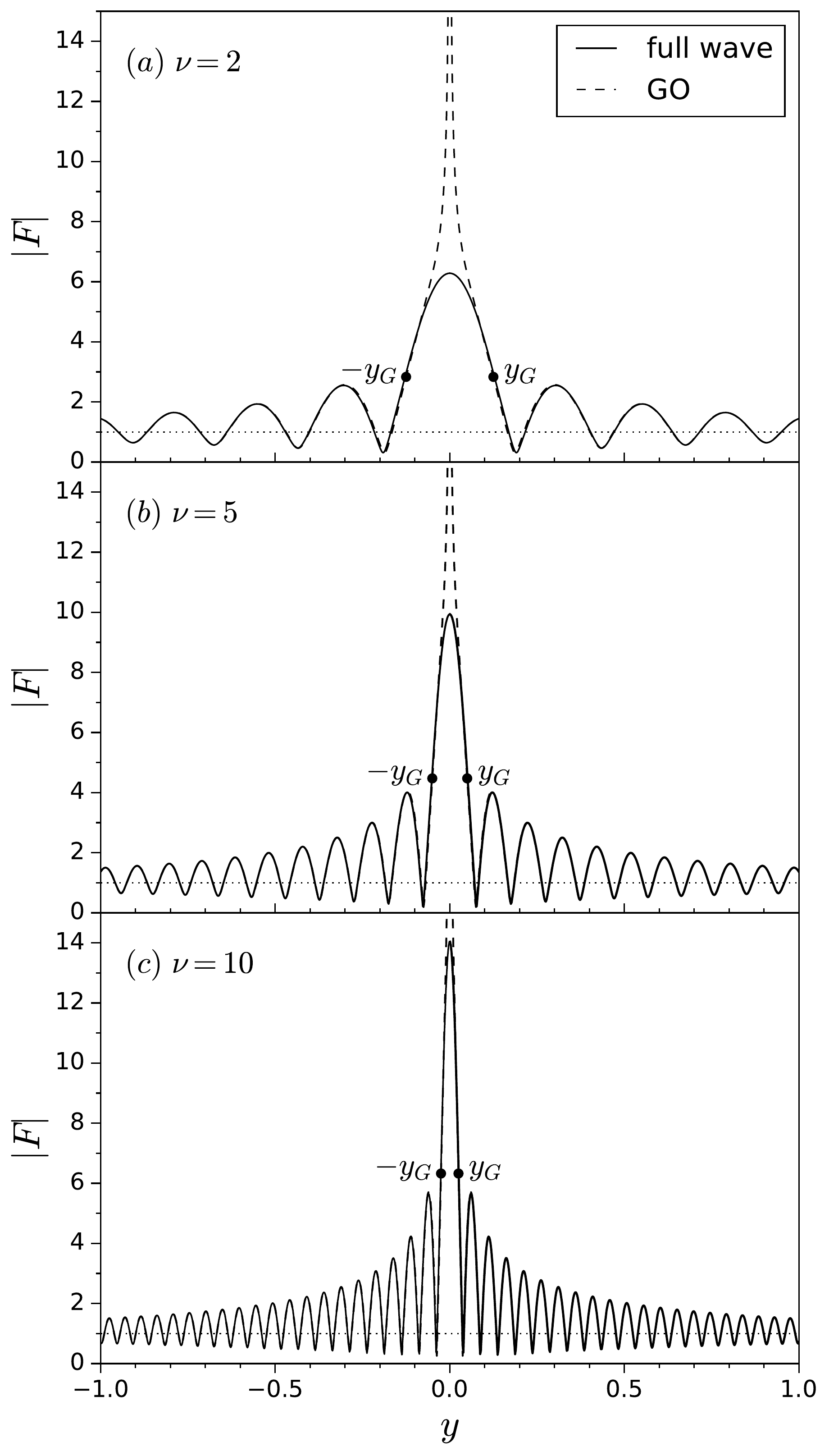}
\caption{Transmission factor $|F|$ vs source location 
$y=\theta_S/\theta_{\rm E}$ for different Fresnel numbers 
$\nu=2; \;5;\; 10$. 
The full-wave solutions \eqref{eq:abs_F} are compared with the GO 
approximations \eqref{F_GO_sin}.}
\label{fig:F-y}
\end{figure}

Fig.~\ref{fig:F-y} shows a comparison of the transmission factor $|F|$ as a 
function of $y$ calculated in the full-wave optics regime and the GO limit, for 
fixed values of $\nu$.
It is seen that the interference pattern is perfectly reproduced by the GO limit 
for $|y|>y_G$, where $y_G = 1/(4\nu)$ is defined similar to $\nu_G$ in 
Sec.~\ref{sec-F-nu}.
The fringe spacing is uniform over the source location $y$ with the characteristic interval
\begin{equation}
\Delta y \approx \frac{1}{2\nu} =\frac{\lambda}{4 R_{\rm S}},
\label{fringe-y}
\end{equation}
i.e., it is proportional to the wavelength $\lambda$ and inversely proportional 
to the mass of the lens. 
It is interesting to note, that the central maximum contains an additional 
information on the topological phase shift between the two images,
since it is wider (see Fig.~\ref{fig:F-y}).
By calculating its width from Eq.~\eqref{eq:hyperbolas} as twice the distance 
between zero and the nearest minimum, we obtain
\begin{equation}
\Delta y_c =  \frac{3}{4\nu} = \Delta y +  \frac{1}{4\nu} =\frac{3\lambda}{8 R_{\rm S}}
\label{fringe-yc}
\end{equation}
where the fringe spacing $\Delta y$ is given by Eq.~\eqref{fringe-y}.
It is seen that the central maximum is wider than the rest of the fringe spacing. 
The additional $1/4\nu$ comes precisely from the Morse phase shift 
$\pi/2$ of the saddle point image.

At $y\to 0$, where the GO approximation fails, one can simplify the full-wave 
solution \eqref{eq:abs_F} by applying the Bessel function approximation 
\cite{deguchi86a}, $\,_1F_1 (\ii\pi\nu;\,1; \,\ii\pi\nu y^2) \approx J_0(2\pi\nu 
y)$. One gets
\begin{equation}
|F(\nu,y)| \approx 
\left( 2 \pi^2 \nu \right)^{1/2} J_0 (2\pi \nu y).
\label{bessel}
\end{equation}
It is also assumed that $e^{-2\pi^2 \nu}$ in \eqref{eq:abs_F} is negligible for 
relevant frequencies 
$\nu \gtrsim 0.3$.
\begin{figure}[t]
\centering
\includegraphics[width=0.7\columnwidth]{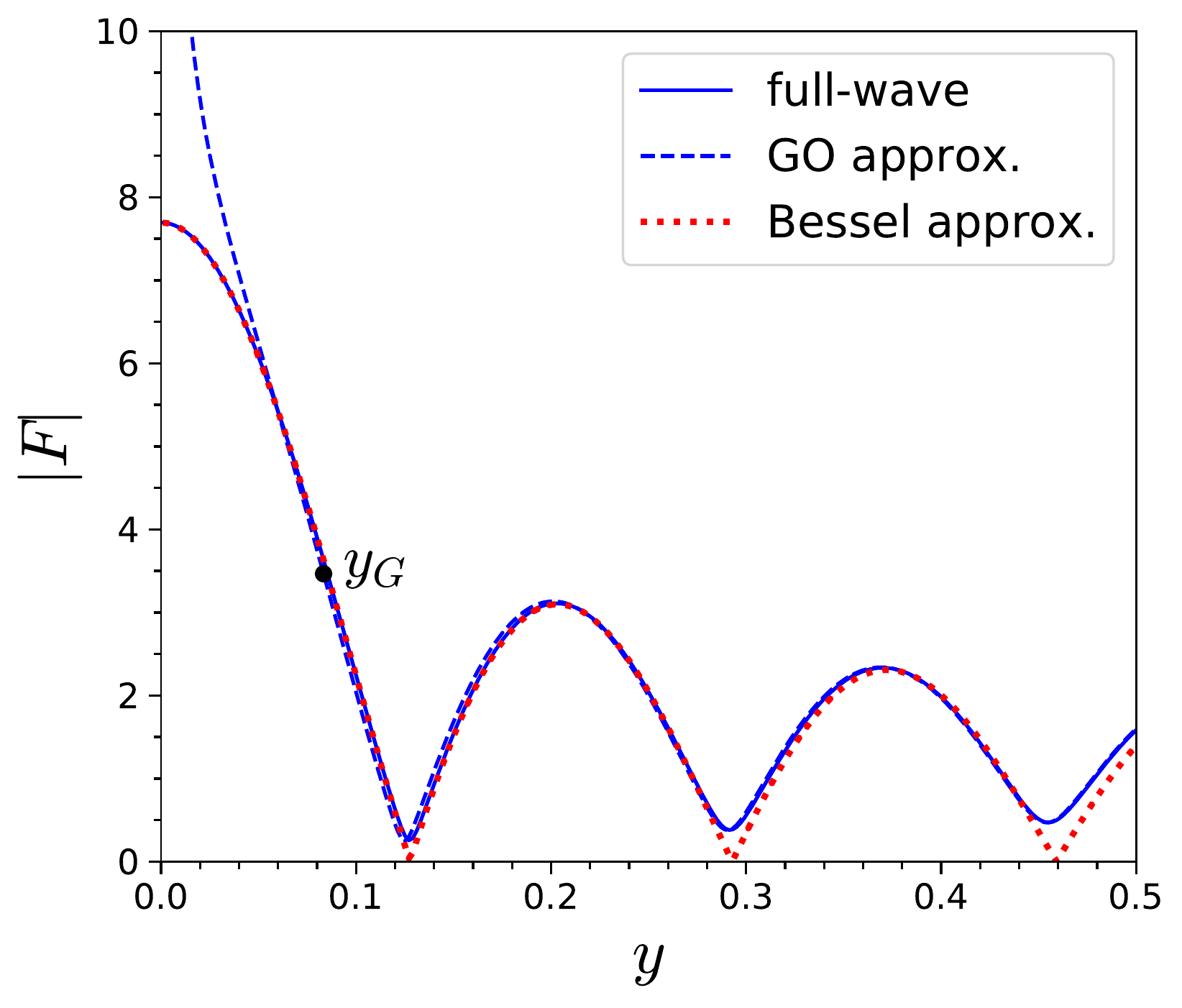}
\caption{Transmission factor $|F|$ vs source location $y$ for $\nu=3$. 
The full-wave solution \eqref{eq:abs_F} is compared with the Bessel approximation  
\eqref{bessel}, valid at small $y$, and the GO approximation \eqref{F_GO_sin}, valid 
for $y>y_G$.}
\label{fig:F-y-Bessel}
\end{figure}
As Fig.~\ref{fig:F-y-Bessel} shows,  Eq.~\eqref{bessel} nicely approximates the 
full-wave solution at low values of $y$ where the GO approximation does not 
hold.

Summarizing this section, the validity of the GO limit can be defined as
\begin{equation}
\nu \,y \gtrsim \frac{1}{4},
\label{eq:GO-valid}
\end{equation}
i.e., for a fixed mass of the lens and source position, the wavelength $\lambda$ 
should satisfy the condition
\begin{equation}
\lambda \lesssim \lambda_G, \qquad 
\lambda_G = 8R_{\rm S} \left(\frac{\theta_S}{\theta_{\rm E}}\right).
\label{eq:GO-valid-lam}
\end{equation}
Thus, the condition for the onset of the GO oscillations depends not only on the ratio between the wavelength $\lambda$ and the Schwarzschild radius $R_{\rm S}$, but also on $y$, the angular location of the source $\theta_S$ (in units of the Einstein angle $\theta_{\rm E}$).

\section{Lensed waveform}

\label{sect-waveform}

In previous sections we have analyzed the transmission factor for a 
monochromatic signal.
Let us now study how a waveform with a given frequency distribution originating 
from a distant GW source is modified by gravitational lensing. 
Since we are interested in wave optics effects, our aim is to investigate how 
the interference between images is imprinted in the lensed waveform. 

Given that we deal with a linear equation for wave propagation, the 
gravitationally lensed waveform $\tilde{h}_L (\omega)$ registered by the 
interferometer in the frequency domain should be the product  \cite{nakamura98,takahashi03}
\begin{equation}
\tilde{h}_L (\omega) = \tilde{h}(\omega) \,F(\omega),
\label{hFh}
\end{equation}
where $\tilde{h}(\omega)$ is the waveform (unlensed) emitted by the source and 
$F(\omega)$ the transmission factor of the lens.

\subsection{Ringdown source}

To simplify analytical treatment, we consider 
a simple waveform with carrier frequency $\omega_0$ modulated by exponentially 
decaying function. This waveform may be associated with the dominant quasi-normal mode of 
the last stage of a binary BH merger, called ringdown 
\cite{flanagan98,berti09}.
The unlensed amplitude for $t>0$ is given by
\begin{equation}
h(t)=h_0 \,e^{-\Gamma t} \cos(\omega_0 t),
\label{h-t}
\end{equation}
where $\Gamma$ is the inverse of the damping time and $h_0$ is the initial 
magnitude.  
For the ringdown model the parameters $\omega_0=2\pi f_0$ and $\Gamma$ are mutually related \cite{berti09}
\begin{equation}
f_0 \simeq 1.207 \times 10^4\, {\rm Hz} 
\,\left(\frac{M_\odot}{M_S}\right),
\quad
\Gamma \simeq 1.496 \cdot f_0
\label{ringdown_qnm2}
\end{equation}
and depend on the mass of the source $M_S$. 
Nevertheless, one could study a more general case as well, when those parameters are independent.
The Fourier transform of the unlensed waveform \eqref{h-t} is
\begin{equation}
\tilde{h}(\omega)=
\int_{-\infty}^{\infty} h(t) \,e^{\ii \omega t} dt =
\frac{\tilde{h}_0 \,(\Gamma-\ii \omega)}{(\Gamma-\ii \omega)^2+\omega_0 ^2},
\label{exp-ul-ft}
\end{equation}
which has a peak close to the carrier frequency $\omega=\omega_0$. The value 
$\tilde{h}_{\rm max}\equiv \tilde{h}(\omega_0)$ is convenient to use for 
normalization 
of the waveform, so that $\tilde{h}_L(\omega)/\tilde{h}_{\rm max}$ will be a 
measure of 
amplification in the frequency domain due to lensing.
As a unit of frequency, it is suitable to introduce the ringdown frequency of the source $\omega_0$. 
Thus, $\tilde{\omega}\equiv \omega/\omega_0$ will be the dimensionless Fourier 
frequency.

For calculations, we use the full-wave transmission factor given by 
Eq.~\eqref{eq:abs_F} for which the variable $\nu$ is translated to the 
dimensionless frequency $\tilde{\omega}$ by $\nu = \nu_0 \,\tilde{\omega}$,
where
\begin{equation}
 \nu_0 = 
 f_0 \, t_M = \frac{2 R_{\rm S}}{\lambda_0}
\end{equation}
is the Fresnel number corresponding to the wavelength $\lambda_0=2\pi c/\omega_0$ of the source.
By its definition, $\nu_0$ is the key wave-optics parameter, which controls the appearance of interference effects.
Since the source frequency $\omega_0$ is determined by the source mass $M_S$, the parameter
$\nu_0$ can also be expressed as the ratio of the masses of the lens and the source 
[substituting the values from \eqref{eq:t_M2} and \eqref{ringdown_qnm2}]:
\begin{equation}
 \nu_0 \simeq  0.24 \, \frac{M}{M_S}.
\end{equation}
We can expect the interference fringe in the lensed waveform  whenever 
the time delay between the images $t_M$ is comparable with the GW period $T_0=1/f_0$. 
Later on, we will specify more precise conditions.
\begin{figure}[t]
\centering
\includegraphics[width=0.48\columnwidth]{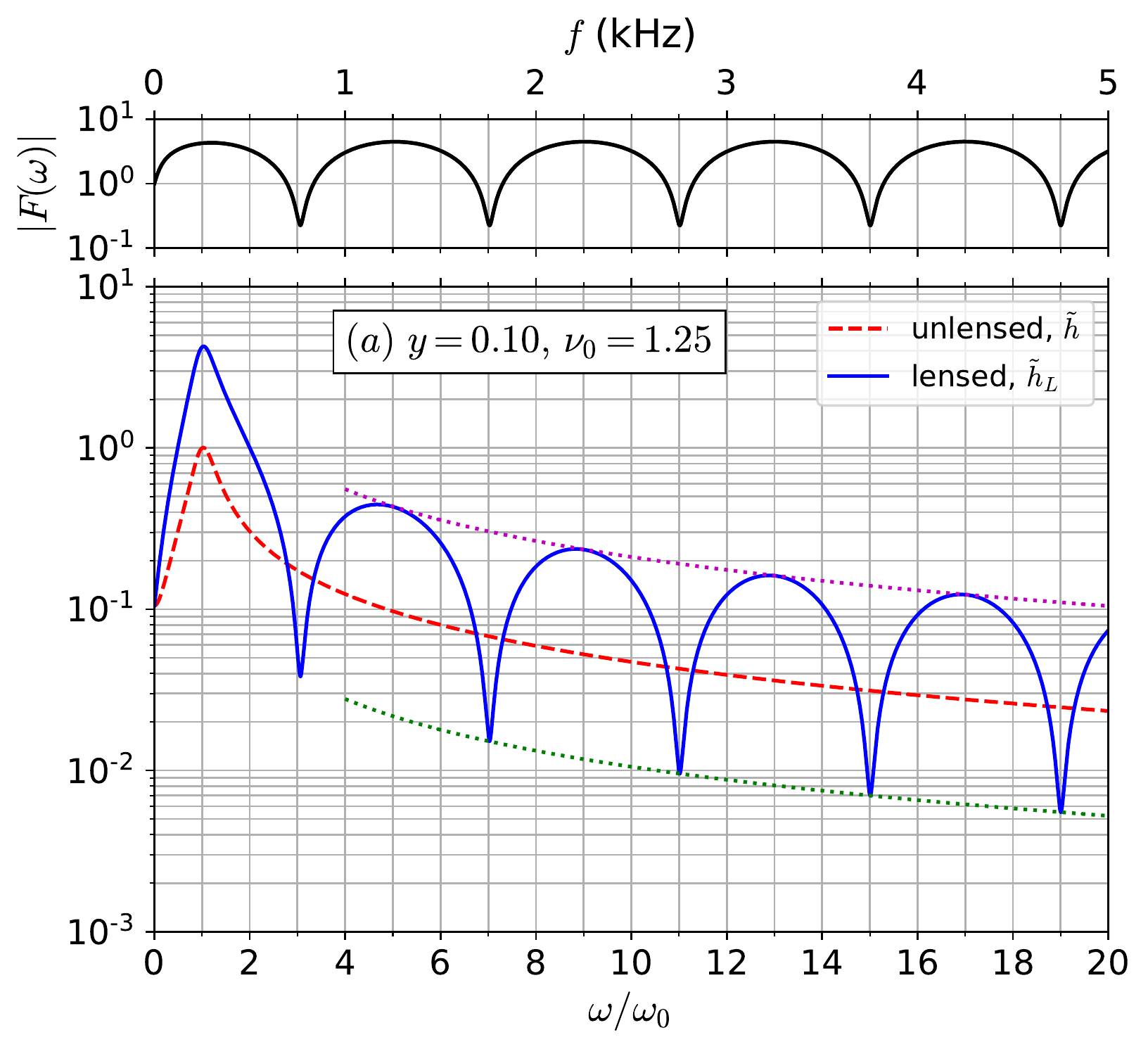}
\includegraphics[width=0.48\columnwidth]{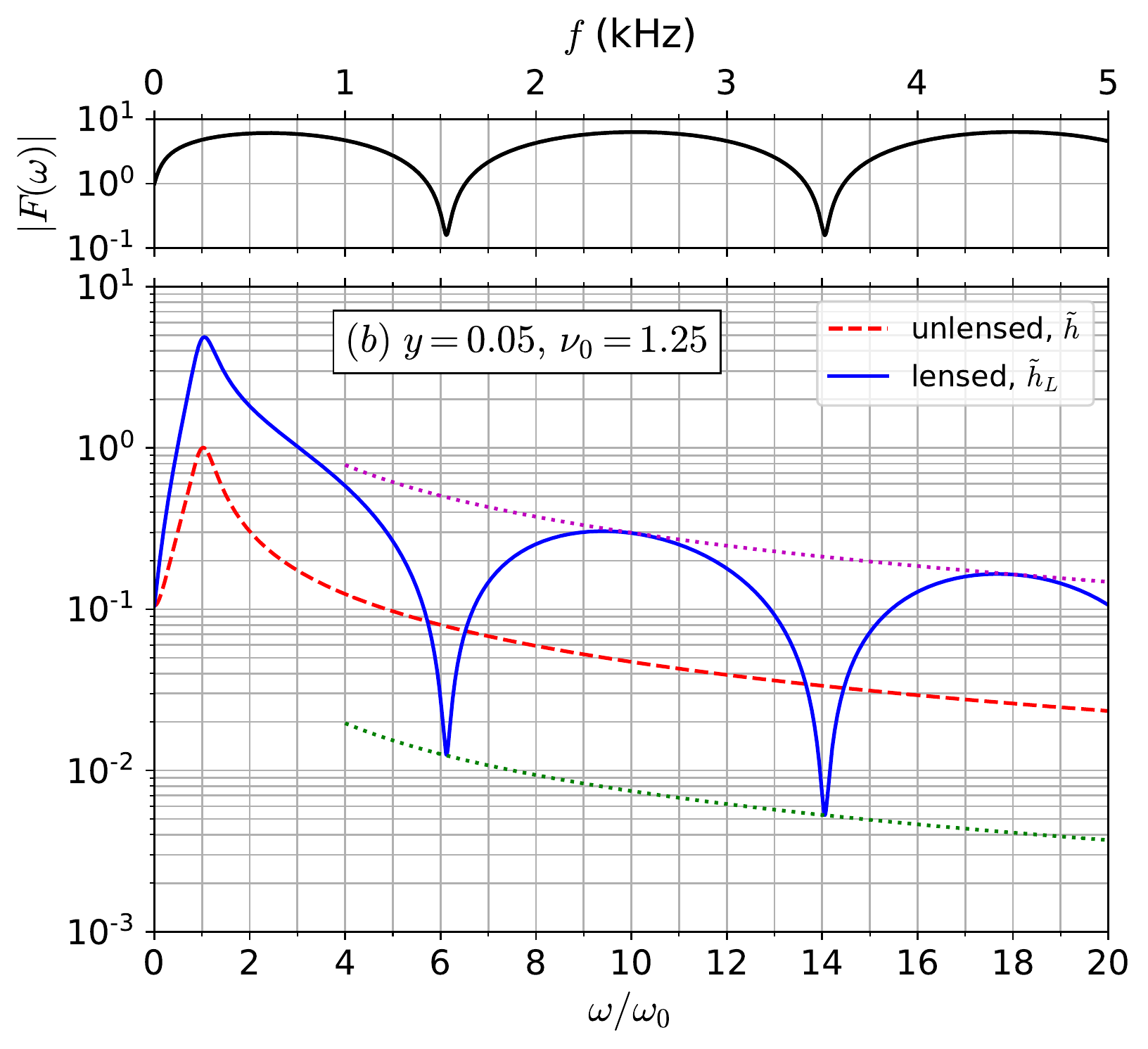}
\caption{Signatures of interference between two images for Fresnel number $\nu_0=5/4$ and two different source locations: (a) $y = 0.1$ and (b) $y=0.05$.
The upper panels of each subfigure show the transmission factor and the lower panels show the strain amplitude in the frequency domain: lensed $\tilde{h}_L(\tilde{\omega})$ vs unlensed $\tilde{h}(\tilde{\omega})$ waveform [Eq.~\eqref{exp-ul-ft}], both normalized to the unlensed peak value. 
The bottom horizontal axis shows the frequency in units of the source frequency $\omega_0$,
while the top axis maps into a physical GW frequency in the LIGO/Virgo band for a specific choice of lens mass 
$M=250 M_\odot$.
The upper envelope line (purple dots) joins the maxima [Eq.~\eqref{eq:max}] and the lower envelope line (green dots) joins the minima [Eq.~\eqref{eq:min}].
}
\label{fig:lensed}
\end{figure}

\subsection{Beating frequencies}

As a first example, consider $\nu_0=5/4$, that corresponds to the lens mass $M\simeq 5M_S$.
For this case, the effect of gravitational lensing on the ringdown waveform is illustrated in Fig.~\ref{fig:lensed}. The interference between two images is manifested as beating fringes in the frequency domain.
The frequencies corresponding to the maxima and minima of the signal can be obtained analytically from the constant-phase lines 
\eqref{eq:hyperbolas}
\begin{equation}
\tilde{\omega}_n = \displaystyle{\frac{1}{2\nu_0 y}\cdot
\begin{cases}
\left( n+\frac{1}{4} \right), \quad \text{maxima,} \\[1mm]
\left( n+\frac{3}{4} \right),  \quad \text{minima},
\end{cases}
}
\label{eq:reson-freq}
\end{equation}
with $n=0,1,2,\dots$. It is easy to see that the values in Fig.~\ref{fig:lensed} are in agreement 
\footnote{except $n=0$ for the first maximum which is outside the region of the GO validity.} 
with the analytical formulas \eqref{eq:reson-freq}. Indeed, for $y = 0.1$ the minima are at $\tilde{\omega}_n = 4n+3=3; 7; 11; 15; \dots$, while for $y = 0.05$  they are at $\tilde{\omega}_n = 2(4n+3)= 6; 14; 22; \dots$. 
In both cases the fringe frequencies are integer multiples of the ringdown frequency $\omega_0$ because of matching 
(taking into account the Morse shift) between $t_M$ and $T_0$ in this case.
Translating into physical units, those frequencies are determined by Eq.~\eqref{eq:hyperbolas_physical}.

\subsection{Beating amplitudes}

The maximum and minimum values of the oscillations of the strain can also be obtained analytically by using Eq.~\eqref{F_GO_sin} for the GO limit or its reduced formula \eqref{cos_approx}. 
For the line joining the maxima (see dotted lines in Fig.~\ref{fig:lensed}) we obtain
\begin{equation}
\tilde{h}_{\rm max}(\omega) = \sqrt{H^+} \,\tilde{h}(\omega), \quad  H^+ 
=\frac{\sqrt{y^2+4}}{y}
\approx \frac{2}{y} + \frac{y}{4},
\label{eq:max}
\end{equation}
while for the minima
\begin{equation}
\tilde{h}_{\rm min}(\omega) =  \sqrt{H^-} \, \tilde{h}(\omega), \quad  H^- 
=\frac{y}{\sqrt{y^2+4}}
\approx \frac{y}{2},
\label{eq:min}
\end{equation}
where the approximations for $H^{\pm}$ are valid under close alignment $y\lesssim 0.5$ and for the frequencies satisfying the GO limit $\tilde{\omega}>1/(4\nu_0 y)$ (equivalent to $\nu>\nu_G$ in Sec.\ref{sec-F-nu}).
The level of amplification of the fringe oscillations, determined as the ratio between the maximum and minimum values, is independent of the frequency 
\begin{equation}
{\cal A} \equiv \frac{\tilde{h}_{\rm max}}{\tilde{h}_{\rm min}}  
= \sqrt{ \frac{H^+}{H^-}} = \sqrt{1+\frac{4}{y^2}}
\approx \frac{2}{y}.
\label{eq:ratio}
\end{equation}
It increases inversely proportional to $y$ near the caustic.

\begin{figure}[t]
\includegraphics[width=0.5\columnwidth]{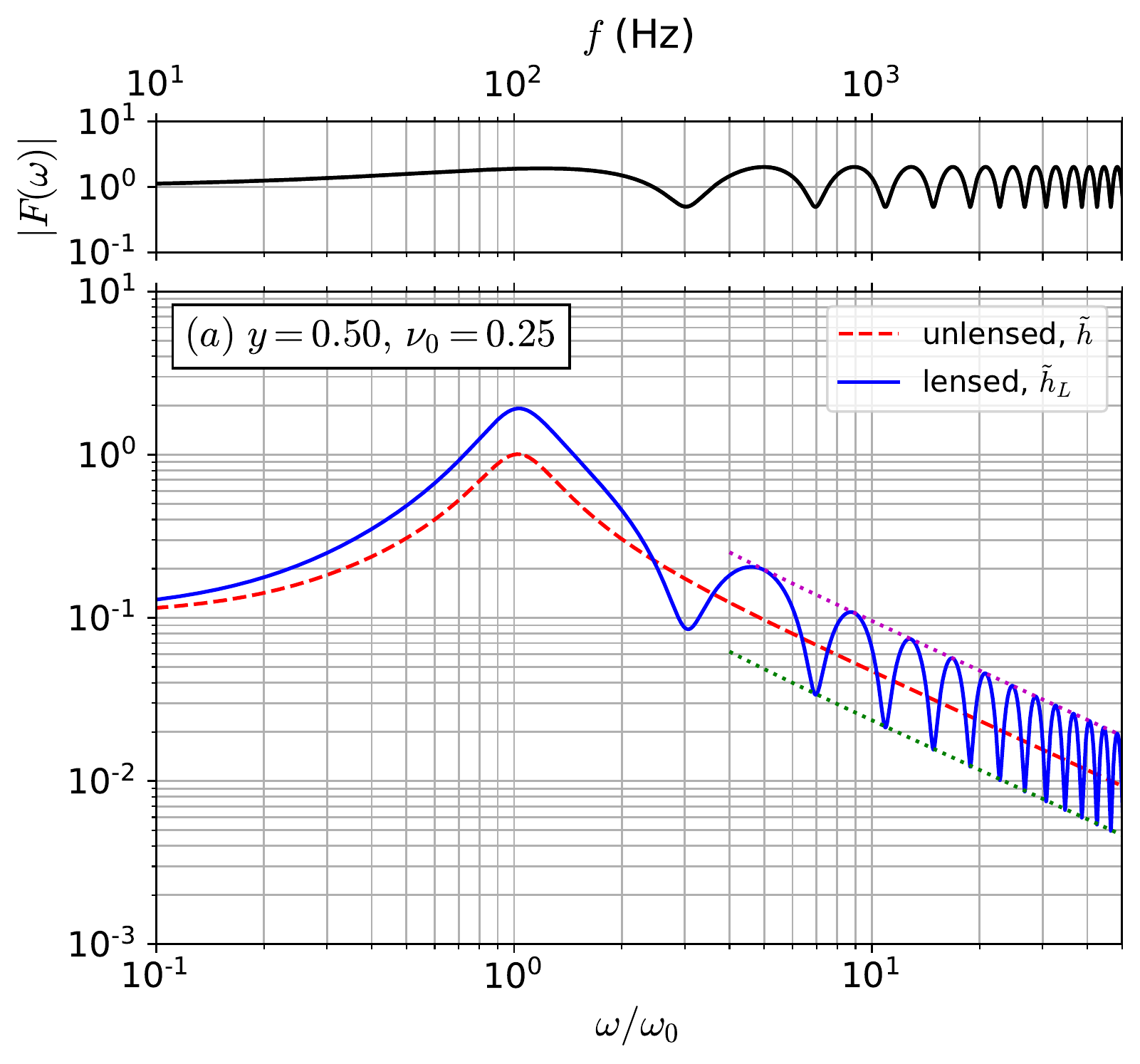}
\includegraphics[width=0.5\columnwidth]{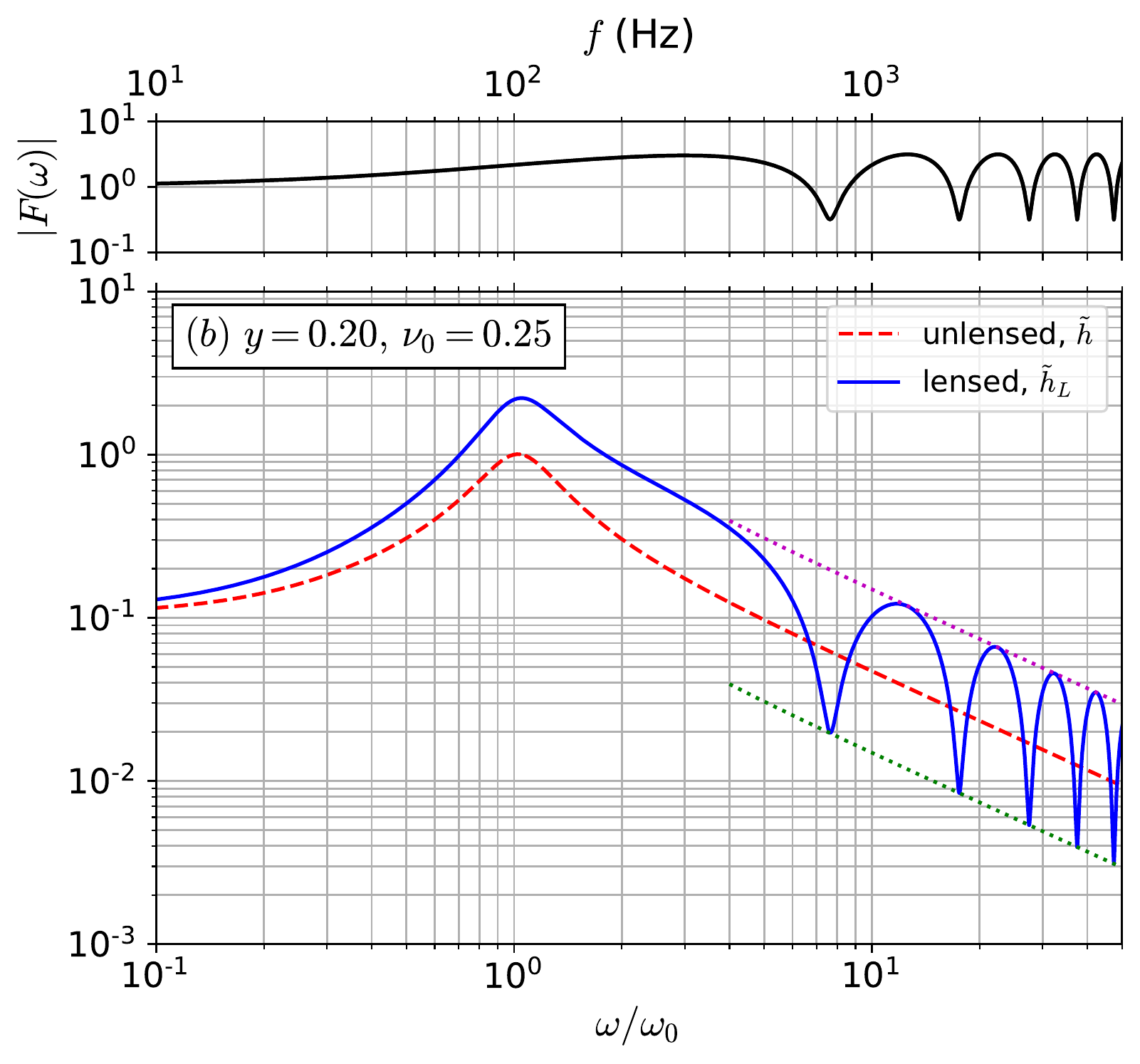}
\includegraphics[width=0.5\columnwidth]{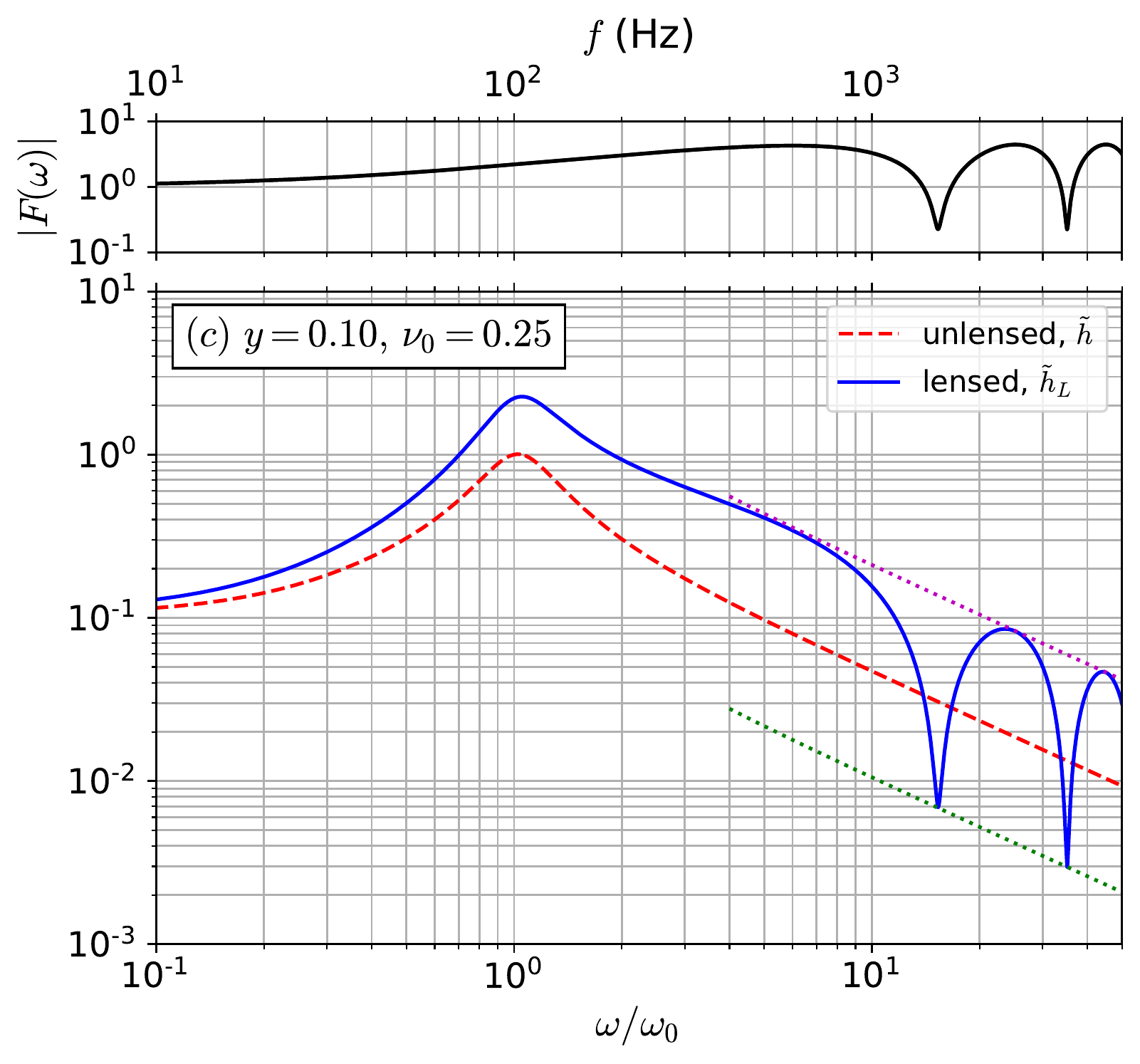}
\includegraphics[width=0.5\columnwidth]{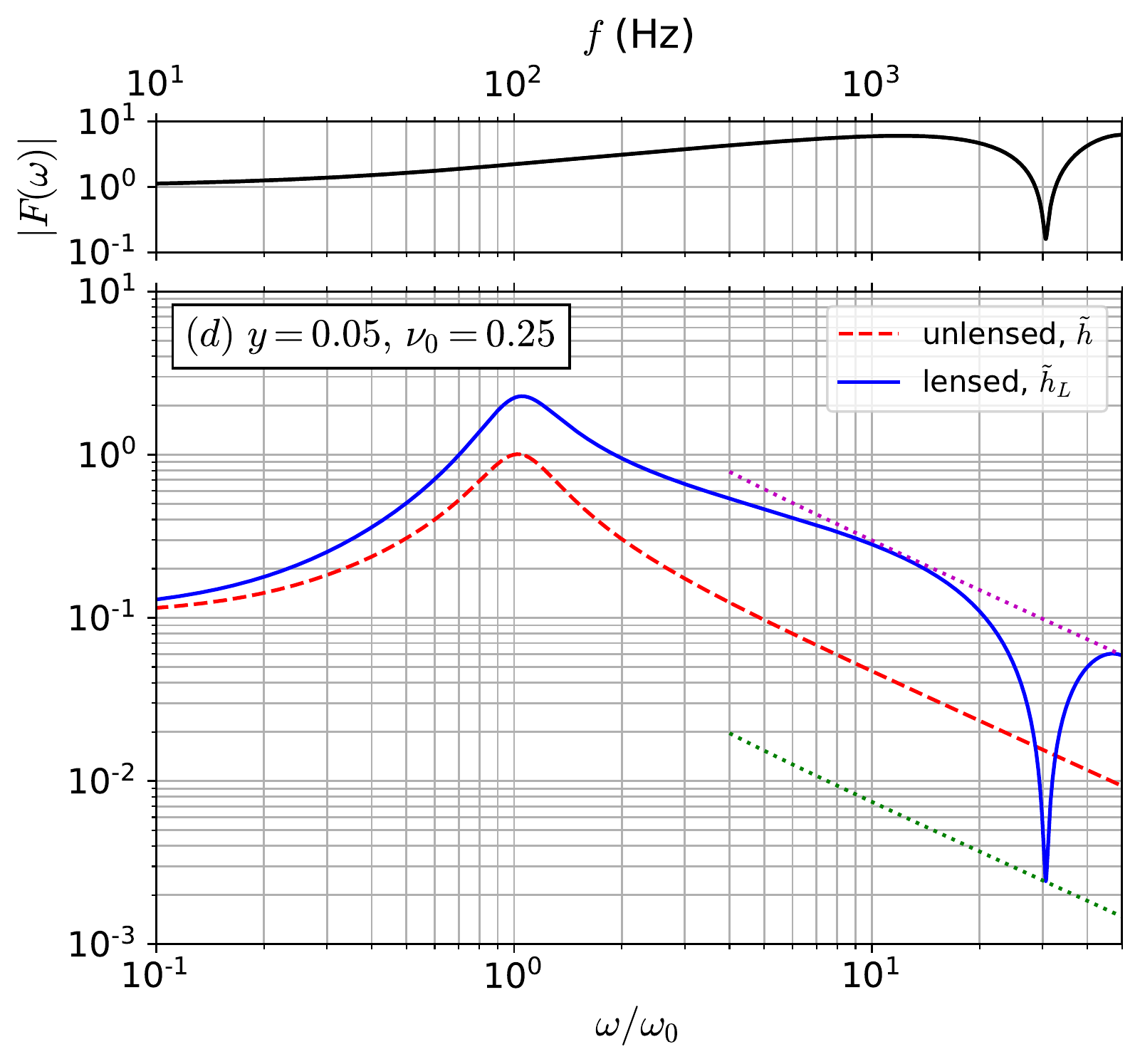}
\caption{Similar to Fig.~\ref{fig:lensed} in log-log scale for Fresnel number $\nu_0=1/4$ and source locations: 
(a) $y = 0.5$; (b) $y=0.2$, (c) $y=0.1$, and (d) $y=0.05$. In all the cases, $\nu_0<\nu_1^+$, so the ringdown principal peak lies at the amplification region of $|F|$ and the GO oscillations appear at higher frequencies. 
The top axis maps into a physical frequency for the lens mass $M=125 M_\odot$.}
\label{fig:lensed-0-25}
\end{figure}

Under the close alignment condition,
$0< y \lesssim 0.5$, one would expect significant amplification, according to Fig.~\ref{fig:F-nu},  for $\nu \gtrsim 0.2$.
As a second example, we consider the lensing for the Fresnel number $\nu_0=1/4$. This value corresponds to the condition that the mass of the lens is approximately equal to the mass of the source, $M\simeq M_S$.
The results are depicted in Fig.~\ref{fig:lensed-0-25} in the log-log scale for different values of $y$ progressively decreasing from $0.5$ to $0.05$ under the close alignment regime approaching the caustic.
The fringe spacing is approximately $\Delta \tilde{\omega} = 1/(2\nu_0y)=2/y$, i.e. inversely proportional to $y$. 
This result can be easily verified from the figure.
The principal ringdown peak is amplified about the factor of 2 and its value is almost independent of $y$, since 
it is located in the ``amplification region" according to Fig.~\ref{fig:F-nu}, which occurs when $\nu_0<\nu_1^+$.
At higher frequencies, the signal gets into the ``GO oscillation region" for which the amplification becomes sensitive
to $y$. 
From Eq.~\eqref{eq:ratio} we obtain the amplification ratio ${\cal A} \approx 4$, $10$, $20$, $40$ 
for $y=0.5$, $0.2$, $0.1$, $0.05$, respectively in the figure.
The closer the source to the caustic, the higher the amplification of the fringe oscillations.
On the other hand, the smaller $y$, the larger the fringe spacing. 

\begin{figure}[t]
\includegraphics[width=0.5\columnwidth]{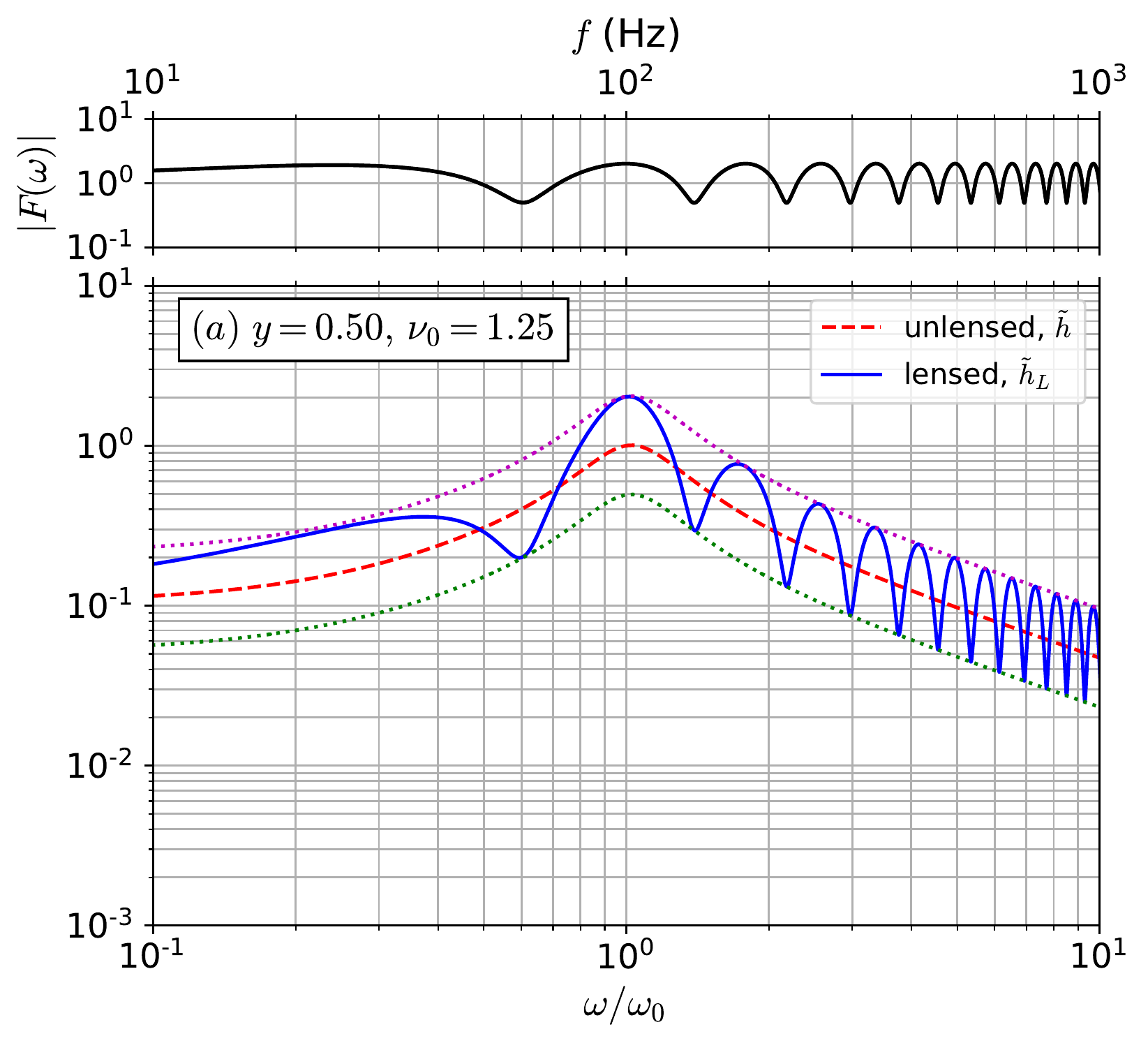}
\includegraphics[width=0.5\columnwidth]{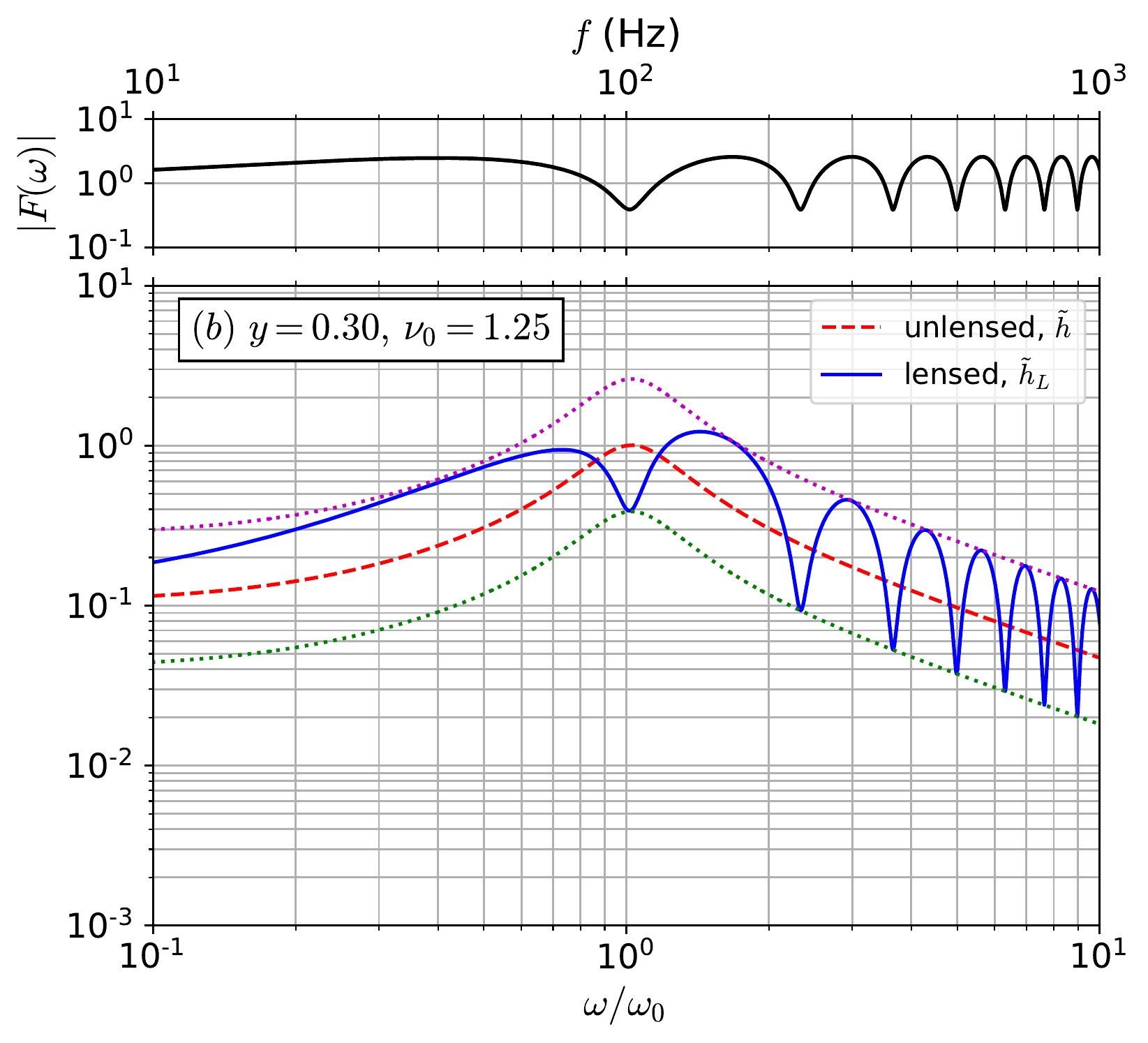}
\caption{Similar to Fig.~\ref{fig:lensed-0-25} for Fresnel number $\nu_0=5/4$ and source locations: 
(a) $y = 0.5$; (b) $y=0.3$, where in both cases $\nu_0>\nu_G$, so the GO oscillations overlap with the ringdown principal peak. 
The upper envelope line (purple dots) joins the maxima [Eq.~\eqref{eq:max}] and the lower envelope line (green dots) joins the minima [Eq.~\eqref{eq:min}]. }
\label{fig:lensed-envelope}
\end{figure}

It can also happen that the principal ringdown peak falls into the GO oscillation region, when $\nu_0>\nu_G$. 
This would eventually occur if the product of the two parameters fulfills the condition $\nu_0 \, y \geq 1/4$. 
For such a case, however, the identification of the unlensed peak position will be more involved, since any of the two interference conditions---constructive or destructive---may happen at the peak frequency, as shown in Fig.~\ref{fig:lensed-envelope}(a),(b).

\subsection{Extraction of the lens mass}

The usual approach to estimate how accurately the parameters, e.g., the mass of the lens, can be extracted from a GW signal
is based on a Bayesian hierarchical analysis using Fisher information matrix, Markov Chain Monte Carlo, or other techniques \cite{takahashi03, cao14,urrutia21}. 
We leave such analysis for future work and 
consider an idealized toy model assuming that the lensing effects are visible in the observed GW waveform.

Suppose we can get independently the fringe spacing $\Delta f$ and the amplification ratio 
${\cal A} =\tilde{h}_{\rm max}/\tilde{h}_{\rm min}$ from the lensed waveform of the observational data.  
Then, by using Eqs.~\eqref{eq:hyperbolas_physical} and \eqref{eq:ratio}, we can infer the time $t_M$ and 
consequently the mass of the lens directly from the measured values:
\begin{align}
t_M &\approx \frac{{\cal A}}{4\Delta f}, \\
\frac{M}{M_{\odot}} &\approx  1.25 \times 10^4 \left(\frac{\rm Hz}{\Delta f}\right) {\cal A}.
\label{eq:mass-extract}
\end{align}
This result suggests the estimation for the lower bound of the lens mass which can be inferred via lensing effect.
As an input parameter we take the high frequency cutoff $f_{max}$ of the detector bandwidth. 
In order to be able to extract information on the lens mass, one needs to observe at least one beating oscillation below the cutoff (roughly the first and the second minimum).
Under this condition, from Eq.~\eqref{eq:hyperbolas_physical} we get
\begin{equation}
\left(\frac{M}{M_{\odot}}\right)_{min}  \simeq  5 \times 10^4 \left(\frac{\rm Hz}{f_{max}}\right) \left(\frac{1}{y}\right).
\end{equation}
For the ground-based interferometers (LIGO, Virgo, KAGRA), if we take the cutoff frequency $f_{max}\simeq  5 \times 10^3\,$Hz \cite{auger-plagnol-17}, 
this gives $(M/M_{\odot})_{min} \simeq 10/y$. Thus, for $y=0.5$, the mass $\gtrsim 20M_{\odot}$ and for $y=0.1$, the mass $\gtrsim 100M_{\odot}$ can in principle be extracted from the lensed waveform oscillations.
As for the mass of the source, since the beating conditions depend only on the lens, it is only necessary that the peak frequency of the source falls into the interferometer bandwidth.

Similar estimations can be done for space-based interferometers (LISA, DECIGO) taking $f_{max}\simeq  1\,$Hz 
\cite{audley17_LISA}.
We obtain $(M/M_{\odot})_{min} \simeq 5\times 10^4/y$. Thus, the lower bound for the mass to be detected would be $M\gtrsim 10^5M_{\odot}$ for $y=0.5$, and $M \gtrsim 5\times 10^5M_{\odot}$ for $y=0.1$. 
The intermediate-mass black holes (IMBHs) and supermassive black holes (SMBHs) \cite{maggiore-18} are possible candidates in this range for detection as a source, as well as a lens via beating pattern in the waveform. 
As previously mentioned, $M$ should be replaced by $M_{zL} = M (1+z_L)$ for the lens at the redshift $z_L$. Similarly, the mass of the source $M_S$ at the redshift $z_S$ should be replaced by $M_S (1+z_S)$.

One can also extract the mass of the source $M_S$ whenever the ringdown principal peak is well separated from the GO oscillations, i.e., when $\nu_0<\nu_1^+$.
In this case the lensed principal peak, even when amplified, will still be at the same frequency $\omega \approx \omega_0$ as the unlensed one (see Figs.~\ref{fig:lensed} and \ref{fig:lensed-0-25}).
The source mass $M_S$ can then be deduced from Eq.~\eqref{ringdown_qnm2}, given the model of one dominant quasi-normal mode is appropriate.
When $\nu_0>\nu_G$, the GO oscillations overlap with the peak (like in Fig.~\ref{fig:lensed-envelope}), so one should locate the position of the peak from the envelope of maxima (or minima).

\section{Summary}
\label{sec-concl}

We have studied the transition from full-wave optics to geometrical optics regimes for gravitational lensing on a compact-mass object (Schwarzschild lens). 
By analyzing the transmission factor in a two-dimensional space of characteristic parameters---the Fresnel number $\nu$ and the source location $y$---we distinguish three physically different regions: diffraction, amplification and geometrical-optics oscillations.
From the point of view of observations, 
two of them---the amplification and GO-oscillations---are of interest.
The latter, in addition, reveals the beating pattern in the waveform for GWs (for EM waves, in most of the cases, the oscillations are washed out due to incoherence and the small size of the wavelength with respect to the size of the source).

Our analysis suggests that the onset of the GO oscillations caused by interference between two images corresponds (with a sufficiently high accuracy) to the condition $\nu y > 1/4$, when we study the close alignment region ($y\lesssim 0.5$), 
which in terms of the time delay between the images can be written in the more general case as $\Delta t_{21}>T/2$, with $T$ being the period of the GW.
The above condition includes the region close to the caustic ($y\to 0$) and it is less restrictive than 
$\Delta t_{21}\gg T$, which is usually considered for the GO limit.
Similarly to light whose intensity is greatest near caustics,  one would expect the highest magnification of GWs---where the lensing effects are likely to be observable---to be close to the caustic as well.
On the other hand, for close alignment the beating pattern in the frequency domain shows universal signatures,
which allow us to infer the (redshifted) mass of the lens from the amplification ratio and the spacing of the fringes [Eq.~\eqref{eq:mass-extract}]. 
While the lens mass can be extracted from the GO oscillations alone, the source location can only be obtained in terms of $y$, i.e., in units of the Einstein angle; there still remains a degeneracy in the distances [Eq.~\eqref{R_E}]. 
Gravitational lensing is a good opportunity to detect the objects which do not emit,
but curve the paths of the incoming waves, causing different parts of the wavefront to interfere, thereby magnifying or distorting the signal.

\acknowledgments

We are grateful to Mark Gieles, Jordi Miralda Escud\'e, Jordi Portell, Tomas Andrade, Ruxandra Bondarescu, Andy Lundgren and the members of the group Virgo--ICCUB for helpful discussions.
H.U.~acknowledges financial  support  from the Institute of Cosmos Sciences of the University of Barcelona (ICCUB).

\bibliography{bib_jcap_pml}

\end{document}